\documentclass[reprint,superscriptaddress,amsmath,floatfix,aps,prd,twocolumn,showkeys]{revtex4-2}
\DeclareUnicodeCharacter{03C9}{$\omega$}
\DeclareUnicodeCharacter{03C3}{$\sigma$}

\usepackage{amssymb}  
\usepackage{graphicx}
\usepackage{svg}
\usepackage{exscale}
\usepackage{textcomp}
\usepackage{enumerate}
\usepackage{amsmath}
\usepackage{amssymb}
\usepackage[compat=1.1.0]{tikz-feynhand} 
\usepackage{multirow}
\usepackage{orcidlink}

\usepackage{hyperref}
\hypersetup{breaklinks=true, colorlinks=true, citecolor=blue}
\usepackage[normalem]{ulem}  
\usepackage{color}

\begin{document}

\title{$f$-mode Oscillations for Hyperons and H-dibaryons in Neutron Stars}
\author{Rajesh Maiti\,\orcidlink{0009-0008-5947-3060}}
\email{rajesh.maiti@iucaa.in}
\affiliation{Inter-University Centre for Astronomy and Astrophysics, Ganeshkind, Pune 411007, India}

\author{Jesper Leong\,\orcidlink{0000-0002-8115-2182}}
\email{jesper.leong@adelaide.edu.au}
\affiliation{CSSM and ARC Centre of Excellence for Dark Matter Particle Physics, Department of Physics, University of Adelaide, 5005, SA, Australia}

\author{Debarati Chatterjee\,\orcidlink{0000-0002-0995-2329}}
\email{debarati@iucaa.in}
\affiliation{Inter-University Centre for Astronomy and Astrophysics, Ganeshkind, Pune 411007, India}

\author{Anthony W. Thomas\,\orcidlink{0000-0003-0026-499X}}
\email{anthony.thomas@adelaide.edu.au}
\affiliation{CSSM and ARC Centre of Excellence for Dark Matter Particle Physics, Department of Physics, Adelaide University, 5005, SA, Australia}


\begin{abstract}
The fundamental ($f$-mode) oscillations of neutron stars are studied within the quark meson coupling model, a relativistic Hartree-Fock theory of dense nuclear matter, which takes into account the self-consistent modification of the valence quark structure of the bound baryons in the associated strong Lorentz scalar mean fields. For the first time, hyperons and H-dibaryons are included, along with the effects of potential additional short-range repulsion within this scheme and their influence on $f$-modes is investigated. Universal relations are studied within the relativistic Cowling approximation and compared against those in the existing literature for potential applications in gravitational wave asteroseismology.
\end{abstract}

\keywords{ neutron star, $f$-mode frequencies, universal relations, equation of state, hyperons, H-dibaryons}

\maketitle


\section{Introduction}
Much of the fascination surrounding neutron stars (NS) stems from the fact that these highly compact astronomical objects provide a natural, extreme environment, unattainable in any terrestrial experiments. Because their core densities can reach many times the saturation density of symmetric nuclear matter ($n_0$), they provide a unique means to constrain the equation of state (EoS) of dense matter~\cite{Lattimer_2021REV,Li:2025uaw}. The most notable constraint from NS observations is that the maximum observed mass can exceed $2M_\odot$, which has not only ruled out many EoS but forced new insights into the composition of the inner core of the NS and the interactions therein. 

It is still unknown whether the core of the heaviest NS contains only nucleons, additional particles including hyperons, or new states of matter, such as Bose Einstein Condensates,  deconfined quark matter or even dark matter~\cite{Annala:2019puf,Whittenbury:2015ziz,Benic:2014jia,Masuda:2012kf,Motta:2022nlj,Agrawal:2009ad,Stone2007,Weber:2006da,Alford:2004pf,Husain:2022bxl}. Besides the mass and radius measurements, several other NS observables such as moment of inertia and compactness may help to distinguish the composition of NS~\cite{Lattimer_2021REV,Breu_2016}. While the binary NS merger event GW170817 is the only gravitational wave (GW) observation to date providing information on the tidal deformability of NS~\cite{LIGOScientific:2018cki, LIGOScientific:2017vwq}, further extractions are expected from planned upgrades to the existing LIGO-Virgo-KAGRA (LVK) detectors, as well as next generation detectors such as the Einstein Telescope (ET) and Cosmic Explorer (CE). These may also allow us to probe NS quasi-normal oscillations, which come in a variety of modes.

Being long-lived, most known NS have had almost all of their heat dissipated, carried off through neutrino emission~\cite{Glendenning:1997wn}. The cold EoS is then subject to $\beta$-equilibrium and the appearance of hyperons is possible, as the Fermi energy of the nucleons reaches the limit where it is energetically more favourable to produce a hyperon with low momentum. One fascinating idea is to invoke a phase transition from nucleonic matter to deconfined quark matter~\cite{Annala:2019puf,Nandi:2017rhy,Li:2021sxb,Ranea-Sandoval:2018bgu,Contrera:2022tqh,Paschalidis:2017qmb}. Other suggestions for the nature of the high density matter include the possibility of a Bose condensate. 
In an interacting system, there is a possibility that the H-dibaryon, a doubly strange dibaryon first postulated by Jaffe~\cite{Jaffe:1976yi} (see also see Ref.~\cite{Mulders:1982da}), may appear. Indeed, recent work including the H-dibaryon has shown compatibility with NS constraints, suggesting that it may exist inside the cores of at least the heaviest NS~\cite{Leong:2025fde,Leong:2025wmh}. 

Although neutron stars with a canonical mass value around $1.4M_\odot$~pose no problems for a wide variety of EoS, it is the massive NS with the highest observed maximum mass, $M_{max}>2.0M_\odot$, which present a significant challenge~\cite{Antoniadis:2013pzd, Nice:2005fi}. The recent NICER mission has provided some of the most precise observations of NS masses and radii to date. In particular, PSR J0740+6620 has a mass of $2.072^{+0.067}_{-0.066}M_\odot$ and a radius of $12.39^{+1.30}_{-0.98}$km~\cite{Miller_2021, Riley:2021pdl, Fonseca:2021wxt, Salmi:2024aum}. The upper limit of $M_{max}$ may be even higher, as suggested by the millisecond pulsar PSR J0952-0607, which has a measured mass of $2.35\pm0.17M_\odot$~\cite{Bassa:2017zpe, Romani:2022jhd}. 

Impressive as these observational results may be, the large uncertainties, especially in the measurements of radii, and model dependencies, do leave one searching for multi-messenger probes of NS. With the detection of gravitational waves from the event GW170817~\cite{LIGOScientific:2017vwq, LIGOScientific:2018cki}, the dimensionless tidal deformability of a $1.4M_\odot$ NS,  $\Lambda_{1.4 M_\odot}=190^{+390}_{-120}$, was extracted. This new NS property is quite sensitive to the radius of a star and therefore provides an independent constraint on it. Note that this is the only GW detection of a binary neutron star merger event to date; an event which had the additional advantage that it was also observed in the electromagnetic spectrum. In addition, one could collate other properties of the star, such as quasi-normal modes (QNM). Still in its infancy, the theoretical study of QNM could help distinguish the interior composition and the interactions which take place at high densities, especially in heavy NS. This would indeed add another layer of constraint~\cite{Pradhan_2021, Pradhan_2022_fullGR}. 

Since non-axisymmetric perturbations lead to the emission of gravitational waves, apart from binary systems, unstable oscillation modes could also produce GW in an isolated NS. The QNM are classified into different categories, $f$-, $p$-, $g$-, and $r$-modes, depending on the restoring forces that bring them back into equilibrium. Pressure and buoyancy act as the restoring forces for the $f$-, $p$-, and $g$-modes, respectively, while the restoring force for the $r$-mode is the Coriolis force~\cite{QNMREV_Kokkotas_1999,Rev_Andersson_2021}. These QNM can be excited by various mechanisms, such as supernova explosions, NS glitches and so on~\cite{Ho_2020,Pradhan_2023_APJ, Wilson_2024}. The GW emitted from these unstable oscillation modes carry information about the NS interior, as their frequencies depend on the internal composition of the star. Among the QNM, we focus on the $f$-mode, as it couples strongly to GW and its expected frequency falls at the upper end of the sensitivity range of the current LVK detectors and well within the range of the next-generation detectors (CE and ET)~\cite{Ho_2020}. Thus, studying the $f$-mode would be very interesting in the context of NS asteroseismology.

In a series of recent works the effects of the internal composition of NS have been examined, alongside systematic investigations of $f$-mode frequencies, by constraining model parameters using nuclear experiments, the properties of finite nuclei at low densities and astrophysical observations at high densities~\cite{Pradhan_2021, Pradhan_2022_fullGR, Montefusco_2024, Jyothilakshmi_2025, Shirke_2024_PRD, Thakur_2024}. These studies employ either the relativistic Cowling approximation or full general relativistic (GR) treatments to calculate the frequencies of the $f$-mode oscillations~\cite{Cowling_1941, Thorne_1967}. The inverse problem of asteroseismology has also been revisited by simulating the observational data expected from next-generation detectors to infer the EoS~\cite{Pradhan_2023_APJ}. 

The study of universal relations (UR) becomes especially important, as these relations correlate various physical properties of NS in an EoS insensitive manner. This allows for the precise determination of one property from the measurement of another, without knowing the exact EoS. In this work, we focus on the relations between the $f$-mode oscillation frequency and the macroscopic properties of neutron stars, such as average density, compactness, tidal deformability and moment of inertia. Andersson and Kokkotas~\cite{Andersson_Kokkotas_PRL_1996} first proposed the study of UR using polytropic EoS. In later works~\cite{Andersson_Kokkotas_MNRAS_1998,Benhar_PRD_2004, Doneva2013,Pradhan_2021, Pradhan_2022_fullGR, Rather_2025, Dey_2025} the empirical relations have been extensively studied for a variety of realistic EoS, including those that incorporate exotic matter components and are compatible with the current constraints. Some of these papers also provided fitting relations within the relativistic Cowling approximation. 

Among the many hadronic theories available, the quark-meson coupling (QMC) model is unique in that the interactions are derived from the self-consistent coupling of the exchanged mesons directly to the valence quarks within a baryon~\cite{Guichon:1987jp,Guichon:1995ue}. As a result of this self-consistency, the Lorentz scalar mean field (carried by the $\sigma$ meson) modifies the valence light quark wave functions in the bound baryons. This change in the internal structure of in-medium baryons leads to density dependent $\sigma$-baryon couplings. These are effectively natural many-body forces associated with changes in the underlying quark-meson dynamics~\cite{Guichon:2004xg,Guichon:2006er}. Although the model has only five parameters, it has shown to be successful in predicting the existence of a heavy NS with hyperons a few years before a $2M_\odot$ NS was experimentally confirmed~\cite{Stone2007}, yielding accurate binding energies and charge radii for the known (over 800) even-even nuclei~\cite{Martinez:2018xep,Martinez:2020ctv}, and predicting binding energies of hypernuclei in reasonable agreement with current experiments~\cite{Stone_2022}. To describe the NS EoS, one should use a theory which is valid at low densities, and at high densities encapsulates the relevant physics, such as subatomic degrees of freedom, where the role of quarks is thought to be important. Although the model has been extensively applied to NS, it has not yet been used to study the $f$-mode oscillations. 

In this work we examine, for the first time, the $f$-modes in the QMC model, with and without the inclusion of H-dibaryons. We emphasise that when modelled in QMC, the interactions of the H arise naturally within the theory itself and require no new parameters. Finally, we also investigate and compare the derived, scaled UR from the resultant $f$-modes. In the next section, section~\ref{sec:method}, the key elements of the QMC formalism and the corresponding EoS are presented. Section~\ref{sec:method} also outlines the calculation of the $f$-modes within the Cowling approximation. The results are presented in section~\ref{sec:results}, along with a detailed discussion, before finishing with some concluding remarks (section \ref{sec:conclusion}).

\section{Methodology}
\label{sec:method}

In NS the strong nuclear force plays a crucial role in generating the pressure necessary to oppose the inward gravitational force. The EoS relates the pressure and energy density of the system. There are a wide variety of possible EoS, depending on the underlying assumptions made within a particular nuclear model. In the following subsection, we provide a description of the most salient features of the QMC model; those which affect the EoS and the $f$-modes. A more detailed account of the model may be found in Ref.~\cite{Guichon:2018uew}. 

\subsection{Features of QMC model}
 In the QMC model the most important forces are carried by the $\sigma$, $\omega$, and $\rho$ mesons. These mesons couple locally to the $u$ and $d$ quarks only, with the $s$ excluded under the OZI rule. What sets this model apart is the coupling of the mesons directly to the quarks. The EoS is calculated in the Hartree-Fock approximation, including the $\pi$ meson.
The in-medium effective mass of the baryon with flavour $f$ is given by 
\begin{equation}
    M^*_f=\frac{N_u\Omega_u+N_d\Omega_d+N_s\Omega_s-z_0}{R_B}+BV+\Delta E_M \, .
    \label{eq:QMCeffM}
\end{equation}
Here, $R_B$ and $B$ are the bag radius and bag pressure appearing in the underlying MIT bag model~\cite{DeGrand:1975cf}. The choice of $R_B$ defines the confinement volume, $V$. Here $z_0$ is the zero-point energy correction resulting from gluon fluctuations across the bag. $\Delta E_M$ describes the spin-dependent one-gluon-exchange interaction, which gives rise to the hyperfine splitting of the masses of hadrons with identical quark content but different spin. The number of $u$, $d$, and $s$ quarks within the baryon is specified by $N_i$. For the $u$ and $d$ quarks the eigenvalues, $\Omega_{u,d}$, depend on the strength of the scalar mean field, as a result of the coupling of the scalar meson to the valence quarks. The hyperfine interaction is also enhanced for this reason. By binding directly to the light quarks, the scalar meson causes a change in the internal dynamics of the baryon, with the biggest change being an enhancement of the lower component of the quark wave function. 

\begin{figure}
 \centering
\includegraphics[width=1\linewidth]{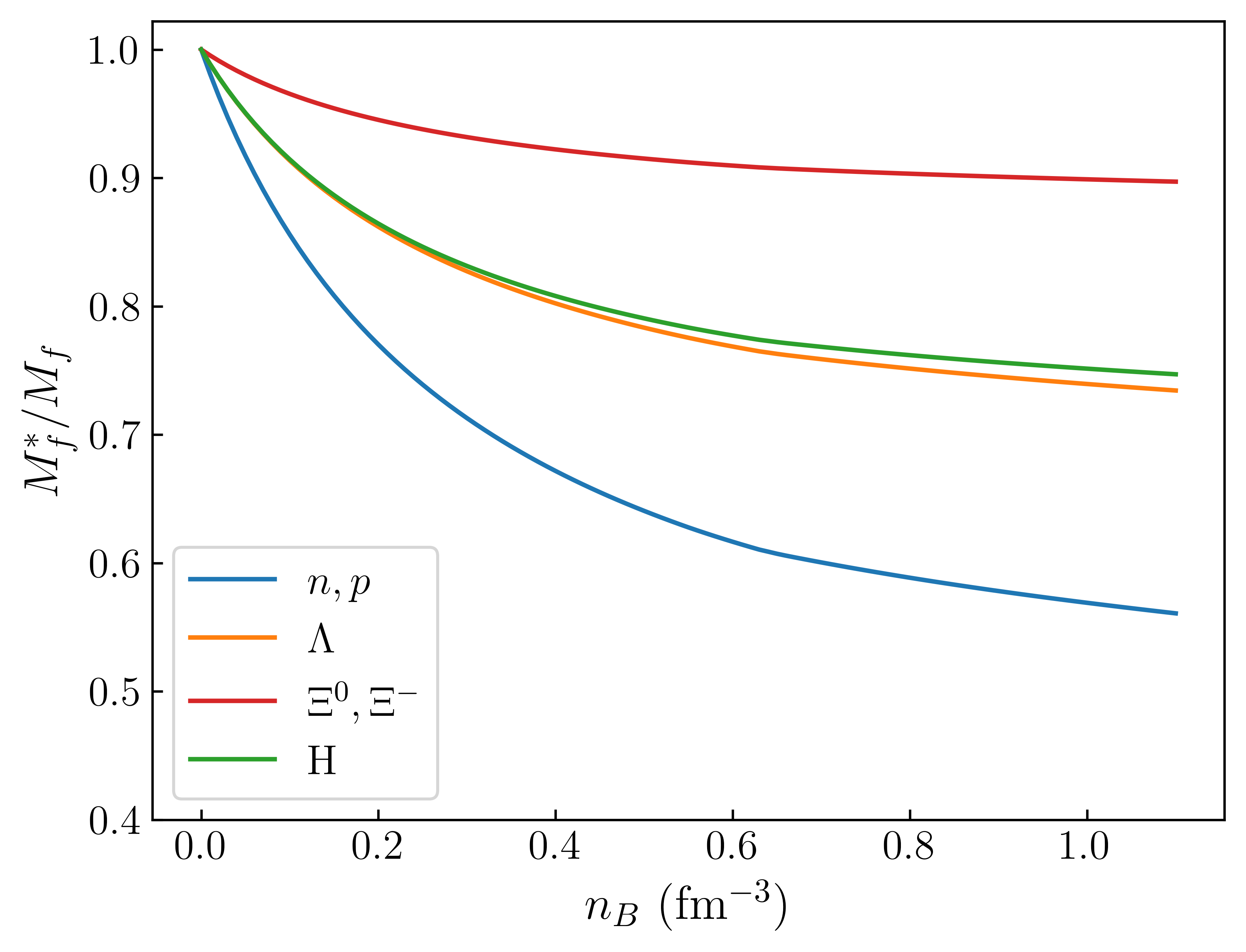}
\caption{The normalised effective masses in QMC Overlap-B with H at $\beta$-equilibrium (cf. Table~\ref{tab:Parameterlist}). The response of the baryon to the scalar field at higher densities is reduced by the effective density dependent couplings in accordance with Eq.~\ref{eq:effM}. }
    \label{fig:effM}
\end{figure}

It is often more practical to express the in-medium mass of the baryon,
Eq.~\ref{eq:QMCeffM}, in terms of the scalar field, as shown for a baryon of flavor $f$ in Eq.~\ref{eq:effM}. The weightings, $w_f^\sigma$ and $\tilde{w}_f^\sigma$, which are calculated in the MIT bag model using a bag radius of $R_N^{free}=1.0$ fm, are baryon flavor dependent and $g_\sigma$ is the $\sigma$-nucleon coupling in free space~\cite{Guichon:2008zz}. $d$ is the scalar polarizability of the nucleon, which encodes the self-consistent effect of the many-body interactions on the $\sigma$-nucleon coupling. Equation~\ref{eq:effM} describes how increases in the scalar field actually result in a decrease of its coupling to a baryon. The effective masses of the nucleons, $\Lambda$, $\Xi^{0,-}$ and the H-dibaryon are shown in Fig.~\ref{fig:effM}. The many-body interaction arising naturally from QMC has 2 major ramifications on the NS EoS; it stiffens the EoS and changes the hyperon thresholds. 
\begin{equation}
    M^*_f=M_f-w_f^\sigma g_\sigma\sigma + \frac{d}{2}\tilde{w}_f^\sigma (g_\sigma \sigma)^2 \, .
    \label{eq:effM}
\end{equation}

The energy of a baryon with flavor $f$ and momentum $\vec{k}$ is given in Eq.~\ref{eq:QMCE}. The $\omega$ coupling is $g_\omega^f=(1+S/3)g_\omega$, with $g_\omega$ the nucleon coupling strength and $S$ the strangeness quantum number. The third component of the isospin of baryon $f$, $I_m^f$, carries the weighting of the $\rho$ coupling to the nucleon, or hyperon, and as such $g_\rho^f=g_\rho$; the nucleon coupling strength. Equation~\ref{eq:QMCE} shows that the vector fields simply shift the energy of the baryon. No additional parameters are necessary when including the hyperons and there is no freedom to vary the in-medium hyperon interactions~\cite{Tsushima:1997cu}. The single particle potentials derived from the underlying model have been shown to yield reasonable agreement with available observations of hypernuclei~\cite{Stone_2022}. Other models typically treat the hyperon potential in Eq.~\ref{eq:QMCE} as a variable, adjusting additional parameters to describe the hyperon  
interactions~\cite{Providencia:2018ywl, Pradhan_2021}. 
\begin{equation}
    E_f=\sqrt{{M^*_f}^2+\vec{k}^2}+g^f_\omega \omega+I_m^f g^f_\rho \rho
    \label{eq:QMCE}
\end{equation}

The energy density shown in Eq.~\ref{eq:Edens} contains terms from the baryons ($\mathcal{H}_B$) and the mesons ($\mathcal{V}_\sigma$, $\mathcal{V}_\omega$, $\mathcal{V}_\rho$). The full expressions, which include the Fock terms, may be found in Ref.~\cite{Leong:2023yma}. The Fock exchange terms arise from fluctuations of the mean-fields and are necessary to incorporate the anti-symmetrization of the baryons. Once the Fock terms are included, the hyperon chemical potentials increase a little~\cite{Stone2007}. This increases the maximum mass of the NS, with a minor change to the radius~\cite{Miyatsu:2011bc, Krein:1998vc, Guichon:2023iev}. In Eq.~\ref{eq:Edens}, $\mathcal{V}_\pi$ denotes the long-range pion exchange, which provides small but meaningful contributions through the corresponding Fock terms~\cite{Stone2007}. 

The final term, $\mathcal{H}_O$, is the overlap energy. A heavy NS is estimated to have a core density between $4-10$ $n_0$, and at these extreme densities baryons are expected to be squeezed together and begin to overlap. At this point the combination of the Pauli-Exclusion Principle, and the increased energy associated with hidden colour configurations in a multi-quark environment are expected to generate additional repulsion at very short distances~\cite{Harvey:1984rq,Bashkanov:2013cla,Brodsky:2011vgv}. This region may also reveal physics beyond the Standard Model~\cite{Berryman:2021jjt}. With this in mind,  high-density repulsion (HDR) has been modelled in the QMC framework using either the overlap term ($\mathcal{H}_O$)~\cite{Leong:2023yma} or the excluded volume effect (EVE)~\cite{Leong:2023lmw, Panda:2002iu, Aguirre:2002xr}. In EVE, the finite size of the baryons leads to a finite volume which no other baryons can enter~\cite{Rischke:1991ke}.
\begin{equation}
    \epsilon_B=\frac{\left < \mathcal{H}_B +\mathcal{V}_\sigma +\mathcal{V}_\omega +\mathcal{V}_\rho +\mathcal{V}_\pi + \mathcal{H}_O \right >}{V} \label{eq:Edens}
\end{equation}

\subsection{The H-dibaryon}
\label{sec:h-dibaryon}
In the 1970s, Jaffe proposed the existence of a deeply bound H-dibaryon, a spinless and isospinless hexaquark with composition $uuddss$~\cite{Jaffe:1976yi}. This led to a number of experiments searching for this particle, with negative results~\cite{Gal:2024nbr,Ejiri:1989rs,Dover:1988hg, Iijima:1992pp, BNLE836:1997dwi, Takahashi:2001nm, Yoon:2007aq}. The NAGARA event at J-PARC ($^{6}_{\Lambda\Lambda}He$) provided the most stringent constraint on the rest mass of H-dibaryon, $M_H>2224$ MeV~\cite{Takahashi:2001nm}. The absence of a deeply bound H is consistent with the analysis of Mulders and Thomas~\cite{Mulders:1982da}, who showed that the larger size of the 6-quark bag would reduce the pion self-energy, leading to a mass around the mass of two $\Lambda$ hyperons. Lattice QCD calculations from the HAL~\cite{INOUE201228} and NPLQCD~\cite{NPLQCD:2012mex} collaborations, found a bound H close to the $\Lambda$-$\Lambda$ threshold but with heavy $u,d$ quarks. This motivated studies of the extrapolation to the physical $u$ and $d$ masses, with Shanahan \textit{et al.}~\cite{Shanahan:2013yta} finding that if it is indeed an exotic 6-quark state the H is unbound by about $26 \pm 11$ MeV. Here we use the nominal value $M_H=2 M_\Lambda+26\, = \, 2258$ MeV. However, given the uncertainties surrounding this particle, we also consider masses roughly 20 MeV higher and lower.

The Lagrangian density for the H dibaryon in the QMC model is
\begin{equation}
    \label{eq:LagH}
    \mathcal{L}_H=\mathcal{D}^*_\mu H^* \mathcal{D}^\mu H- {M^*_H}^2H^*H \, ,
\end{equation}
where $H$ represents the H-particle wave function and $\mathcal{D}^\mu= \partial^\mu-ig^H_\omega \omega^\mu$.  
In the QMC model the coupling to the $\omega$ field is $g_\omega^H=\frac{4}{3}g_\omega$, which is the same as Ref.~\cite{Glendenning:1998wp}, since  simply counts the number of non-strange quarks. Unlike other studies, the effective mass of the H-dibaryon, $M^*_H$, follows from Eq.~\ref{eq:effM} and contains the non-linear scalar polarizability term. This is given as
\begin{eqnarray}
    &M^*_H=M_H-2\times[0.6672+0.04638 R_H-0.0022 {R_H}^2] g_\sigma \sigma \nonumber \\
    &+2\times[0.00146+0.0691R_H-0.00862 {R_H}^2](g_\sigma \sigma)^2. \, . 
    \label{eq:effMH}
\end{eqnarray}
The weightings in Eq.~\ref{eq:effMH} are derived from knowing that the H has twice the quark composition of the $\Lambda$ and take into account the increased size of the six-quark MIT bag~\cite{Mulders:1982da}. Thus the coupling of the H to the $\sigma$ filed can be written in terms of $g_\sigma$, the nucleon coupling. Here we set $R_H=1.2 R_N^{free}$~\cite{Mulders:1982da}. A more detailed explanation of this may be found in Ref.~\cite{Leong:2025fde}. As for the hyperons, the model involves no additional parameters to describe the interactions involving the H-dibaryon, except its unknown rest. 

From Eq.~\ref{eq:LagH}, the energy density associated with the H is
\begin{equation}
\label{eq:Henergy}
    \epsilon_H=2 {M_H^*}^2 H^*H= M^*_H n_H,
\end{equation}
and the chemical potential
\begin{equation}
    \mu_H=M^*_H+g_\omega^H\omega \, . \label{eq:chemH}
\end{equation}
The discussion of the modifications to the mean-fields are omitted here but may be found in Ref.~\cite{Leong:2025fde}. Note that because the H has zero spin and isospin it does not couple to the $\rho$-field, and there are no Fock terms (and by extension, no pion contributions). Following from Eq.~\ref{eq:Henergy}, the H does not directly contribute to the pressure. But in an interacting H system the pressure does change because of the modification of the $\sigma$ and $\omega$ fields. We do note that when applying the high density overlap contribution the H does directly contribute to the pressure~\cite{Leong:2025fde}.

\subsection{QMC parametrisation}
\label{sec:QMC_parametrisation}
There are five parameters needed to define the QMC model. The meson-quark couplings may be represented by the meson-nucleon couplings in free space, $g_\sigma$, $g_\omega$, and $g_\rho$. These are chosen to reproduce the nuclear matter properties (NMP); saturation density ($n_0=0.14-0.17$ fm$^{-3}$), binding energy per nucleon ($E/A=-15.8\pm1.2$ MeV) and symmetry energy ($S_0=30\pm2$ MeV). Relativistic theories tend to produce an incompressibility higher than the preferred range,  $K_\infty=240-260$ MeV~\cite{Stone:2014wza}. The inclusion of a cubic self-interaction term in the $\sigma$ field has been found to reduce the incompressibility. It is governed by the coefficient $\lambda_3$, and based on calculations of the energies of giant monopole resonances, this is estimated to lie in the range $\lambda_3 \in 0.02-0.05$ fm$^{-1}$~\cite{Martinez:2018xep}. The final QMC parameter is $m_\sigma$, which is generally taken to be in the range $m_\sigma=500-800$ MeV.

 We have chosen to use the same parametrisations found in Refs.~\cite{Leong:2023yma} and \cite{Leong:2023lmw}, which modelled the QMC EoS with heavy NS constraints including the overlap interaction and EVE correction, respectively. It should be noted that both of these studies optimised the EoS under the assumption that the star contains hyperons, and did not selectively set the parameters to include the H. In Ref.~\cite{Leong:2023yma}, the overlap parameters used were $E_0=5500$ MeV and $b=0.5$ fm. In Ref.~\cite{Leong:2023lmw} the EVE hardcore radius was chosen to be $r=0.45$ fm. In both overlap and EVE the HDR is dependent on the total number density, which includes the H-particle. The inclusion of the H uses the same QMC parameters  found in Refs.~\cite{Leong:2025fde} and \cite{Leong:2025wmh}. As here, both of those studies set $M_H\in [ 2247, 2258, 2269 ]$ MeV in order to test the sensitivity of the conclusions to the unknown mass. In particular, here we are concerned with the sensitivity of the $f$-mode frequency to the value of $M_H$. Unless otherwise stated, all calculations are performed using $M_H=2258$ MeV. Table~\ref{tab:Parameterlist} summarises the QMC parametrizations used here to explore their effects within the Cowling approximation for the $f$-mode frequency.

\begin{table*}[ht]
    \centering
    \begin{tabular}{|c|ccccccc|}
    \hline
        Label & $m_\sigma$ (MeV) & $\lambda_3$ (fm$^{-1}$) & $n_0$ (fm$^{-3}$) & E/A (MeV) & $S_0$ (MeV) & $K_\infty$ (MeV) & $L_0$ (MeV) \\ \hline 
         & \multicolumn{7}{c|}{Standard (QMC)}  \\ \hline
        QMC-A & 700 & 0.0  & 0.16 & -15.8 & 30 & 289 & 52 \\ 
         QMC-B & 700 & 0.02 & 0.16 & -15.8 & 30 & 257 & 53 \\ 
        QMC-C & 500 & 0.05 & 0.15 & -15.0 & 30 & 225 & 58 \\ \hline
        & \multicolumn{7}{c|}{High Density Repulsion (HDR)} \\ \hline
        Overlap-A & 700 & 0.0  & 0.16 & -15.8 & 30 & 293 & 52 \\ 
        Overlap-B & 700 & 0.02 & 0.16 & -15.8 & 30 & 260 & 53 \\ 
        EVE-C & 500 & 0.05 & 0.15 & -15.0 & 30 & 257 & 61 \\ 
        \hline
    \end{tabular}
    \caption{The QMC coupling strengths are fitted to reproduce $n_0$, $E/A$, and $S_0$, once $m_\sigma$ and $\lambda_3$ are chosen. The incompressibility ($K_\infty$) and slope of the symmetry energy ($L_0$) at saturation are computed once the parameters are fixed. The six EoS are divided into the standard QMC model and QMC with HDR. The application of HDR uses either the overlap phenomenology, found in Ref.~\cite{Leong:2023yma}, or the EVE correction, detailed in Ref.~\cite{Leong:2023lmw}. A further six EoS include the H-particle, employing the same parametrisations as those given above.}
    \label{tab:Parameterlist}
\end{table*}

\subsection{NS Composition, EoS and Macroscopic NS Properties}

The NS composition is constructed by invoking $\beta$-equilibrium. The energy density is minimised whilst conserving electrical charge and baryon number. Under $\beta$-equilibrium, it seems  that hyperons must form within a heavy NS since its central density is thought to be as high as $4-10$ $n_0$; which hyperons and at what abundances are subject to the hyperon potentials of the nuclear theory being used. Because of the density dependence of the hyperfine interaction, the $\Sigma$ and $\Delta$ baryons do not appear in the final QMC NS EoS~\cite{Motta:2019ywl,Stone2007}. If it does exist, the H-particle will only appear if, according to Eq.~\ref{eq:chemH}, the relevant chemical potentials satisfy the condition $\mu_H=2\mu_N$. 

\begin{figure*}[ht]
    \centering
    \includegraphics[width=1\linewidth]{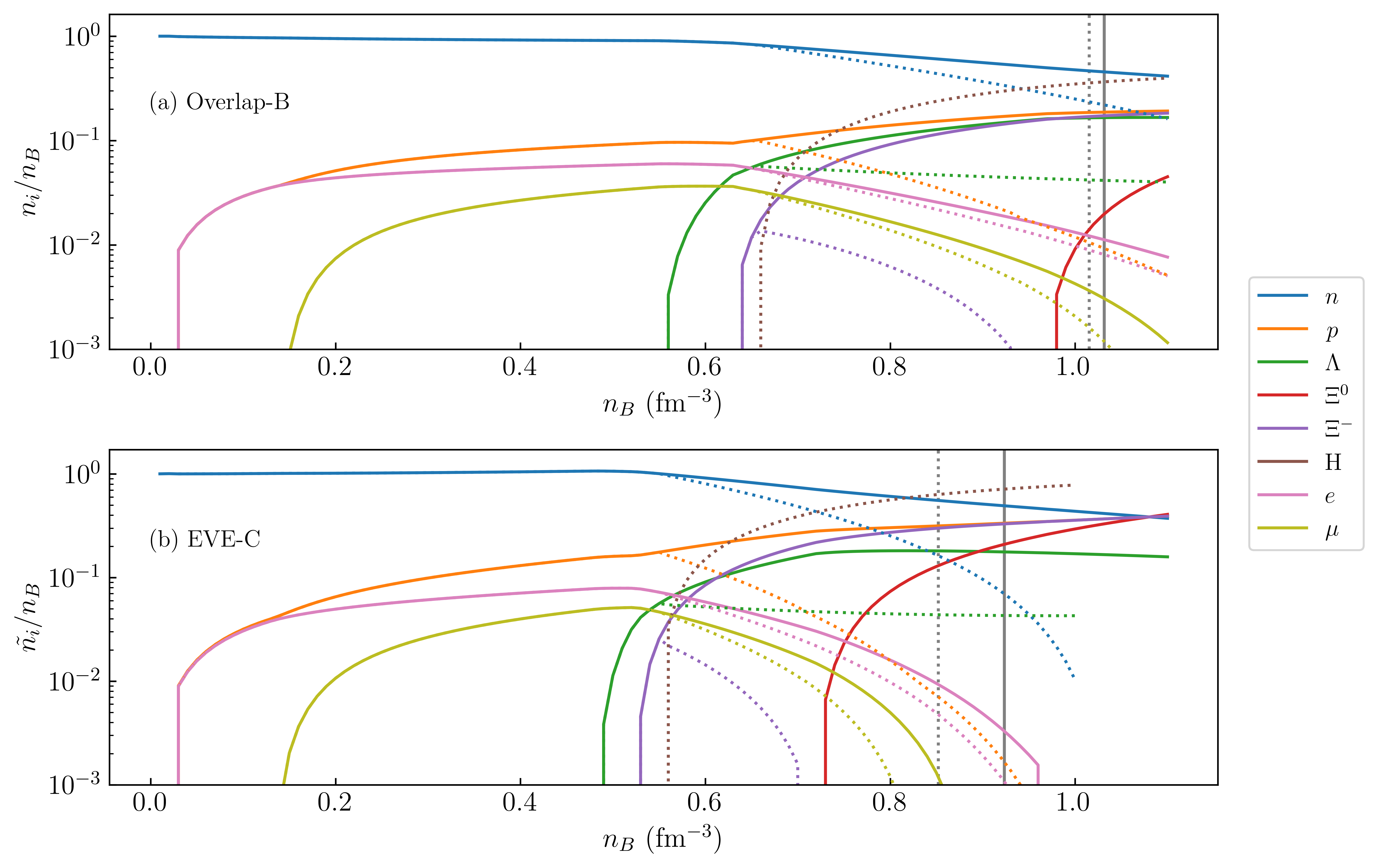}
    \caption{Particle fractions with increasing density have been displayed for the Overlap-B (top-panel) and the EVE-C (bottom-panel) parameter sets, respectively, as discussed in the Table~\ref{tab:Parameterlist}. The solid (dotted) lines represent the system without (with) H-dibaryons. The vertical solid (dotted) grey lines indicate the central density corresponding to the maximum mass configuration without (with) H-dibaryons. }
    \label{fig:particle_fraction}
\end{figure*}

The particle fractions in QMC are displayed in 
Fig.~\ref{fig:particle_fraction}, comparing the cases with and without the H-dibaryon. 
Here we have chosen to display only the Overlap-B and EVE-C parameter sets, both of which produce an acceptable incompressibility because of the non-zero $\lambda_3$, as well as and satisfying the heavy mass NS constraint. Overlap-A has $\lambda_3=0$ fm$^{-1}$, which is the only difference from Overlap-B. This causes the incompressibility to be undesirably high as shown in Table~\ref{tab:Parameterlist}. There is a small difference in the hyperon and H thresholds but ostensibly the particles maintain the same order of appearance. The effects of $\lambda_3$ on the particle fractions were discussed in Ref.~\cite{Leong:2023yma}.

Because $M_H>2M_\Lambda$, the H particle always appears after the $\Lambda$. Clearly the onset of the H is dependent on the choice of $M_H$ and its interactions, both of which are differ from Ref.~\cite{Glendenning:1998wp}. 
Mantziris {\em et al.}~\cite{Mantziris:2020xwi} explored a range of potentials for a different di-baryon, the $d^*(2380)$, interacting in a nucleon only medium. Celi \textit{et al.} extended this work to include hyperons, exploring the couplings of $d^*$ between the non-interacting case and twice the nucleon coupling strength~\cite{Celi:2023gtj}. Within the QMC model the chemical potential of the $d^*(2380)$ never lies below that of two neutrons and so it can never appear in our NS. 
As seen in Fig.~\ref{fig:particle_fraction}, when the H is excluded at $\beta$-equilibrium, the $\Xi^0$ appears in the NS. However, when the H is included, the $\Xi^0$ is no longer present. Another consequence is that particulate species which are already present in the EoS, have their abundances decreased. When the H is present, the H quickly  dominates and is likely to be the most abundant species, even more so than neutrons, for NS near the maximum mass. This is not just a peculiarity of QMC, as both Glendenning and Schaffner-Bielich find that the H is favoured in high mass stars~\cite{Glendenning:1998wp}. 

The compatibility of the H particle with NS observations is also determined by the degree of softening the H-particle has on the EoS. The H, as previously discussed, does not contribute directly to the pressure. A stable star containing the H is only possible in an interacting system, because of the indirect modification of the pressure associated with the changes in the $\sigma$ and $\omega$ fields, which the H generates.

\begin{figure*}[!h]
    \centering
    \includegraphics[width=1.0\linewidth]{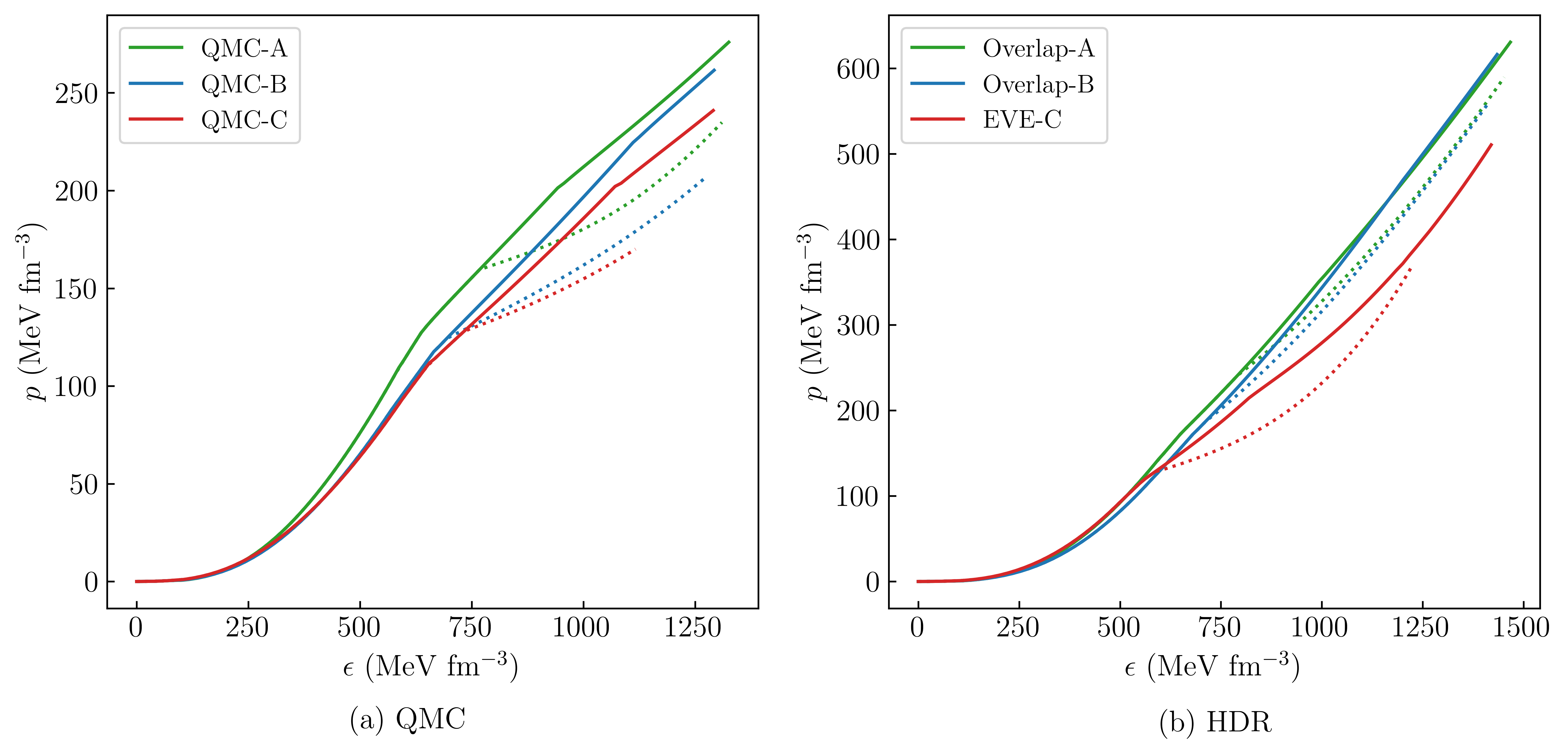}
    \caption{The EoS for different QMC parametrisations used in this work for both a) QMC and b) HDR. The solid lines correspond to the case where nucleons and hyperons are included, while the dotted lines indicate the effect of adding the H-dibaryon.}
    \label{fig:eos}
\end{figure*}

Figure~\ref{fig:eos} shows all the EoS presented in Table~\ref{tab:Parameterlist}. At low densities the softness of the EoS can be related to the incompressibility. Higher values of $K_\infty$, found when $\lambda_3=0.0$ fm$^{-1}$, do generate stiffer EoS~\cite{Leong:2023yma}.
In Fig.~\ref{fig:eos} (b), we show that the application of the HDR mitigates the softening of the EoS caused by the H. The HDR modelling used here assumes that at higher densities, there are additional repulsive interactions caused by a combination of the Pauli-Exclusion Principle and a multiquark environment. The overlap and EVE corrections are flavor independent, include the hexaquark, and tend to  dominate at high densities.

\begin{figure*}[!h]
    \centering
    \includegraphics[width=1.0\linewidth]{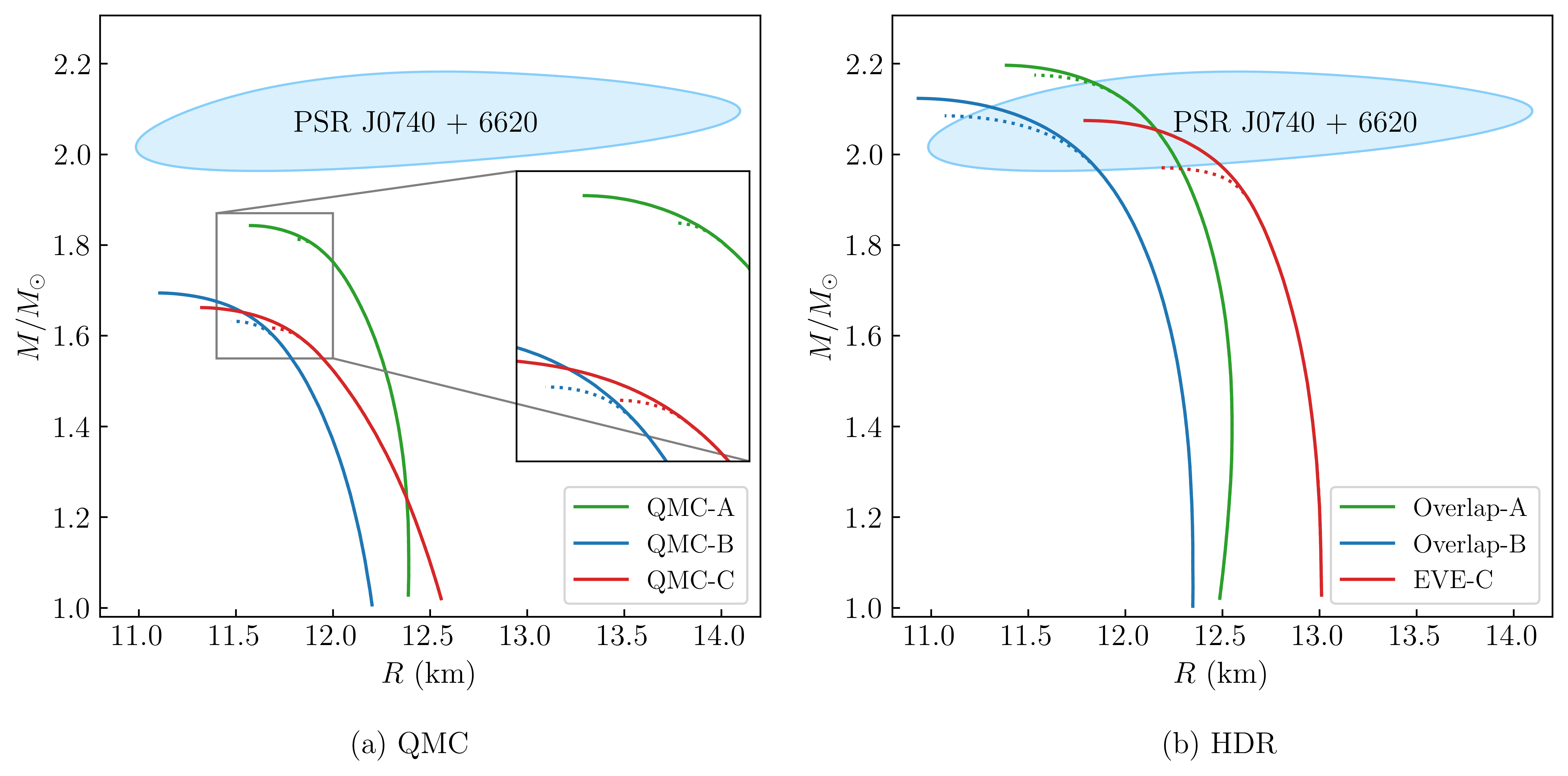}
    \caption{The mass-radius relations for the EoS parametrisations shown in Fig.~\ref{fig:eos}. In a) QMC, the H particle almost immediately collapses the neutron star, making it difficult to distinguish from the solid line. This region has been magnified for the reader. b) shows the HDR EoS. The $1\sigma$ contour of the NICER mass-radius measurement of pulsar ``PSR J$0740+6620$" has been displayed~\cite{Salmi:2024aum}.}
    \label{fig:MR}
\end{figure*}

The mass-radius curves displayed in Fig.~\ref{fig:MR} (b) extend beyond $2.0 M_\odot$, which is only possible with these QMC parametrisations (with low incompressibility) because of the HDR. We first note that the H-particle is only present in heavy mass stars. 
From Fig.~\ref{fig:particle_fraction}, we find that the hyperons appear when $n_B>3n_0$, which corresponds to stars with mass at least $1.8M_\odot$. 
Furthermore, as mentioned above, the result of both flavor independent overlap and EVE is to allow a star containing the H-partilce to be stable for longer, as evident by the reduction in radius from the point for the H-particle's appearance. 

Using HDR also has an effect at low mass. A comparison between Figs.~\ref{fig:MR} (a) QMC and \ref{fig:MR} (b) HDR, shows that the latter has a larger radius for low mass stars.  
As no hyperons or H-particles are present in low mass stars, the tidal deformation is not presented here; it is identical to the results presented in Refs.~\cite{Leong:2023yma, Leong:2023lmw}. 


\subsection{Calculation of $f$-modes}
\label{sec:formalism_fmode}
Non-radial oscillation modes of NS have been a subject of extensive study for several decades. In the non-relativistic framework, Cowling was the first to propose a method for analysing these modes~\cite{Cowling_1941}, while the full GR approach was later developed by Thorne and Campollataro~\cite{Thorne_1967}. In the full GR treatment, one must account for metric perturbations when solving the perturbed fluid equations, making the process computationally intensive. However, these metric perturbations are neglected within the relativistic Cowling approximation, which simplifies the analysis. This approximation has been widely used in the literature for studying $f$-modes~\cite{Sotani_2011,Doneva2013,Flores2017,Sandoval2018,Pradhan_2021}. It has been shown that the $f$-mode frequency computed within the relativistic Cowling approximation is $20-30\%$ higher than the frequency obtained in the full GR treatment~\cite{Yoshida1997,Pradhan_2022_fullGR}. Since the metric perturbations are neglected, this approximation does not allow for the calculation of the damping timescale of the QNM and hence the frequencies obtained in this approximation are real.

In this work, we adopt the relativistic Cowling approximation, and hence the metric for the spherically symmetric background is given by,
\begin{equation}
    ds^2 = -e^{2\Phi(r)} dt^2 + e^{2\lambda(r)}dr^2 + r^2d\theta^2 + r^2\sin^2\theta d\phi^2 \, .
\end{equation}
The equations governing the fluid oscillations can be obtained by perturbing the conservation equation of the energy-momentum tensor. The Lagrangian displacement vector that characterizes the motion of the fluid element is described as
\begin{equation}
    \zeta^i = \left(e^{-\lambda(r)}W(r), -V(r)\partial_\theta, -V(r)\sin^{-2}\theta \partial_\phi \right)r^{-2} Y_{lm}(\theta, \phi)~,
\end{equation}
where $Y_{lm}$ are the spherical harmonics. Assuming that the time dependence is harmonic, the two functions $W(r)$ and $V(r)$ satisfy the following coupled equations~\cite{Sotani_2011,Pradhan_2021}:
\begin{align}
    \label{eq:f1}
    \frac{dW(r)}{dr} = \,&\frac{d\epsilon}{dp} \left[ \omega^2 r^2 e^{\lambda (r) - 2\Phi(r)}V(r) + \frac{d\Phi(r)}{dr} W(r) \right] \nonumber \\
    & -l(l+1)e^{\lambda (r)} V(r)~,\\
    \label{eq:f2}
    \frac{dV(r)}{dr} =\; & 2\frac{d\Phi(r)}{dr} V(r) - \frac{1}{r^2} e^{\lambda (r)} W(r)~,
\end{align}
where
\begin{equation}
    \frac{d\Phi(r)}{dr} = -\left[\frac{1}{\epsilon(r) + p(r)}\right]\frac{dp}{dr}~.
\end{equation}

The asymptotic behaviour of the functions $W(r)$ and $V(r)$ near the center is given by
\begin{equation}
    W(r) = A r^{l+1};\; V(r) = -\frac{A}{l}r^l \, .
\end{equation}
Here, $A$ is an arbitrary constant. For a given EoS,  $p=p(\epsilon)$, one must solve the above differential equations together with the TOV equations to determine the eigenfrequencies ($\omega$) that satisfy the following boundary condition 
at the surface, for each value of $l$,
\begin{equation}
    \omega^2e^{\lambda(R) - 2\Phi(R)}V(R) + \frac{1}{R^2}\frac{d\Phi(r)}{dr}\Big|_{r=R} W(R) = 0 \, ,
\end{equation}
where $R$ is the radius of the NS.
The $f$-mode frequencies are calculated for $l=2$, for NS with masses between $1.0M_\odot$ and the maximum mass in all analyses. 

\section{Results}
\label{sec:results}
The results are divided into 2 parts. First, we present the $f$-modes for all QMC EoS derived in section~\ref{sec:QMC_parametrisation}. We describe how the QMC parameters affect the $f$-modes through the EoS and how the results compare with previous studies which explore hyperonic degrees of freedom. Then, in section~\ref{sec:URs}, we study the UR employing the QMC parametrisations and compare our results with the existing literature. Finally, we discuss the detectability of $f$-modes in section~\ref{sec:detectability}.

\subsection{QMC $f$-mode characteristics}
\begin{figure*}[ht]
    \centering
    \includegraphics[width=1.0\linewidth]{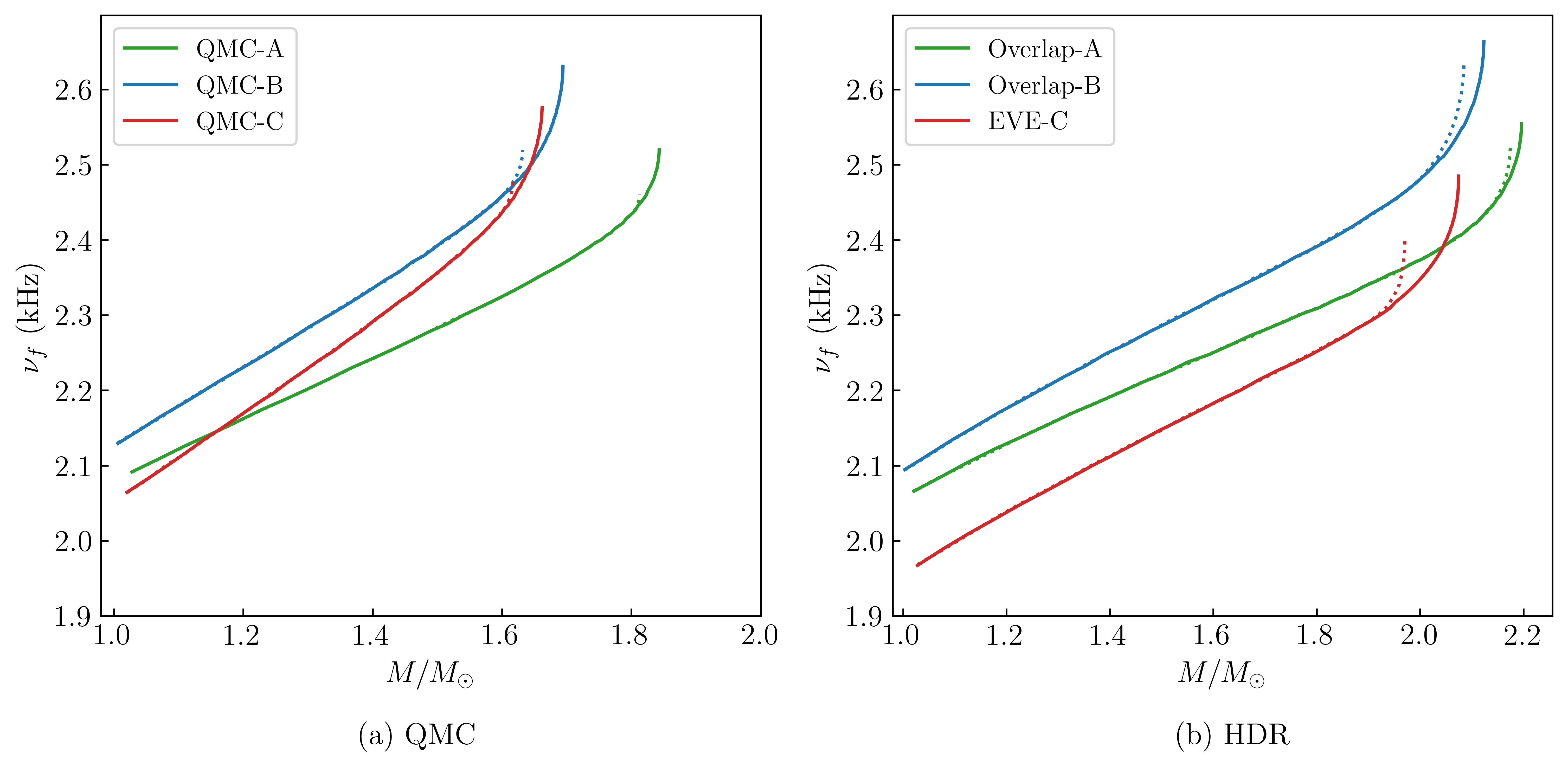}
    \caption{The $f$-mode frequencies (for $l=2$) as a function of NS mass. The solid curves correspond to EoS with both nucleons and hyperons. The dotted curves indicate that nucleons, hyperons and H-dibaryons are included.}
    \label{fig:fmodes}
\end{figure*}

The $f$-modes for the 12 EoS are displayed as a function of NS mass in Fig.~\ref{fig:fmodes}. There is a wide spread of $f$-mode frequencies ($\nu_f$). In the low NS mass region the frequencies lie between $1.97-2.13$ kHz, while at high NS mass they vary between $2.40-2.66$ kHz. At low masses, the higher $f$-mode frequencies are attributed to softer EoS at low density, which are more heavily linked to how the couplings are set to reproduce the NMP and the resulting incompressibility. For reference, the baryonic densities for a $1.0M_\odot$ NS lie between $0.28-0.34$ fm$^{-3}$, when only nucleons are present. When the self-interaction of the $\sigma$-field is included (i.e., $\lambda_3 \neq 0$ fm$^{-1}$), the energy density includes more attraction, generates a softer EoS and the $f$-mode frequencies are found to be higher. In general, the HDR increases the pressure at supra-saturation density and the result is a lower $f$-mode frequency at low NS mass. The type of HDR is also important to consider. The EVE formalism changes the NMP and stiffens the EoS more at low densities than the overlap phenomenology. Compared to the overlap case, the $f$-mode frequencies of the EVE are found to be lower. 

The softening of the EoS has generally been offered as an explanation of why hyperonic EoS show an increase in $\nu_f$, when compared to nucleonic EoS. This is not unique to hyperons, as $\Delta$ baryons show similar effects~\cite{Rather_2025}. Rather \textit{et al.} found that the initial appearance of the $\Delta$ baryons, which in their work occurred in low mass stars, created a softer EoS resulting in a larger frequency when compared to both nucleon only and hyperonic EoS~\cite{Rather_2025}. Hence, one may naively expect that the H-particle would also produce a higher $f$-mode frequency because of the additional degree of freedom. From Fig.~\ref{fig:fmodes}, one can observe that the $f$-mode frequencies for the cases with and without the H-dibaryon differ only for massive neutron stars (with masses $> 1.8M_\odot$ for the HDR EoS parametrizations). This is because the H-dibaryon appears only at very high densities and is therefore confined to the cores of such massive stars. To quantify this difference, we calculate and report the relative frequency difference in Table~\ref{tb:rel_freq_diff} at the maximum masses of the H-containing configurations for the HDR EoS parametrizations that satisfy the observed $2M_\odot$ NS constraint. One can verify that the relative difference in frequency between the cases with and without H is at most $\sim 3\%$ for the EoS parametrizations.

\begin{table}[!ht]
    \centering
     \resizebox{1.0 \linewidth}{!}{
    \begin{tabular}    {|c|c|c|}
    \hline
    EoS parametrizations & {$\nu_f$ } (kHz)  & $|\Delta \nu_f/\nu_f|$ (\%) \\ \hline
   Overlap-A (without H) & 2.48 & \multirow{2}{2em}{1.65}   \\
    
   Overlap-A (with H) & 2.53 &   \\ \hline

   Overlap-B (without H) & 2.55 & \multirow{2}{2em}{3.37}   \\
   Overlap-B (with H) & 2.64 &   \\ \hline

   EVE-C (without H) & 2.33 & \multirow{2}{2em}{3.17}  \\
   EVE-C (with H) & 2.40 &   \\ \hline
    \end{tabular}
    }
    \caption{The relative differences in frequencies (computed using unrounded values) between the cases with and without H are shown at the maximum masses of the corresponding EoS parametrizations with the H particle ($M_{max,H}$). The values of $M_{max,H}$ for Overlap-A, Overlap-B, and EVE-C are  $2.18M_\odot$, $2.09M_\odot$, and $1.97M_\odot$, respectively.} 
    \label{tb:rel_freq_diff}
\end{table}

Our results show that indeed $\nu_f$ does increase upon the appearance of H, in all parametrisations, which is consistent with expectation. However, at maximum mass, NS which contain H-dibaryons do not have an $f$-mode frequency which is larger than those containing just hyperons at their maximum mass. Fig.~\ref{fig:particle_fraction} illustrates how the introduction of the H-particle is essentially in competition with other baryonic species. It is noted that $\Xi^0$ does not appear in the NS core. The H directly modifies the $\sigma$ and $\omega$ fields and, although the H is not directly involved, the $\rho$ and $\pi$ strengths are different because of the changes in abundance of the other particles present. 

There are some cases where the inclusion of the H, although initially softening the EoS, may lead to a stiffer EoS at higher densities. This can be seen from the EVE-C curve in Fig.~\ref{fig:eos}. EVE-C without (solid) and with (dotted) the H-particle start to differ when the H appears. Without the H particle, EVE-C has $\Xi^0$ appearing at around $n_B=0.73$ fm$^{-3}$, corresponding to an energy density of 
830~MeV~fm$^{-3}$, which softens the EoS. The $\Xi^0$ does not appear if the H is included (c.f. 
Fig.~\ref{fig:particle_fraction}) and the pressure rises again at higher densities. This effect is seen in all H and non-H comparisons but is most noticeable for the EVE, because it lowers the hyperon thresholds. 
The $\Xi^0$ contributes less to the $\omega$-field and facilitates additional $\pi$ exchanges, which drives up the $f$-mode frequencies. Furthermore, at maximum mass, NS with the H have larger radii (c.f. Fig.~\ref{fig:MR}). Our conclusion is that the appearance of additional degrees of freedom, normally leading to a reduction in neutron Fermi-pressure, is dependent on the competition of the H with the other baryons. In this case the H replaces the $\Xi^0$. There is an increase in the $f$-mode frequency as the H forms, but the lower mass at which these NS become unstable results in a lower $\nu_f$, when the comparison is made at maximum mass.
\begin{figure*}[ht]
    \centering
    \includegraphics[width=1.0\linewidth]{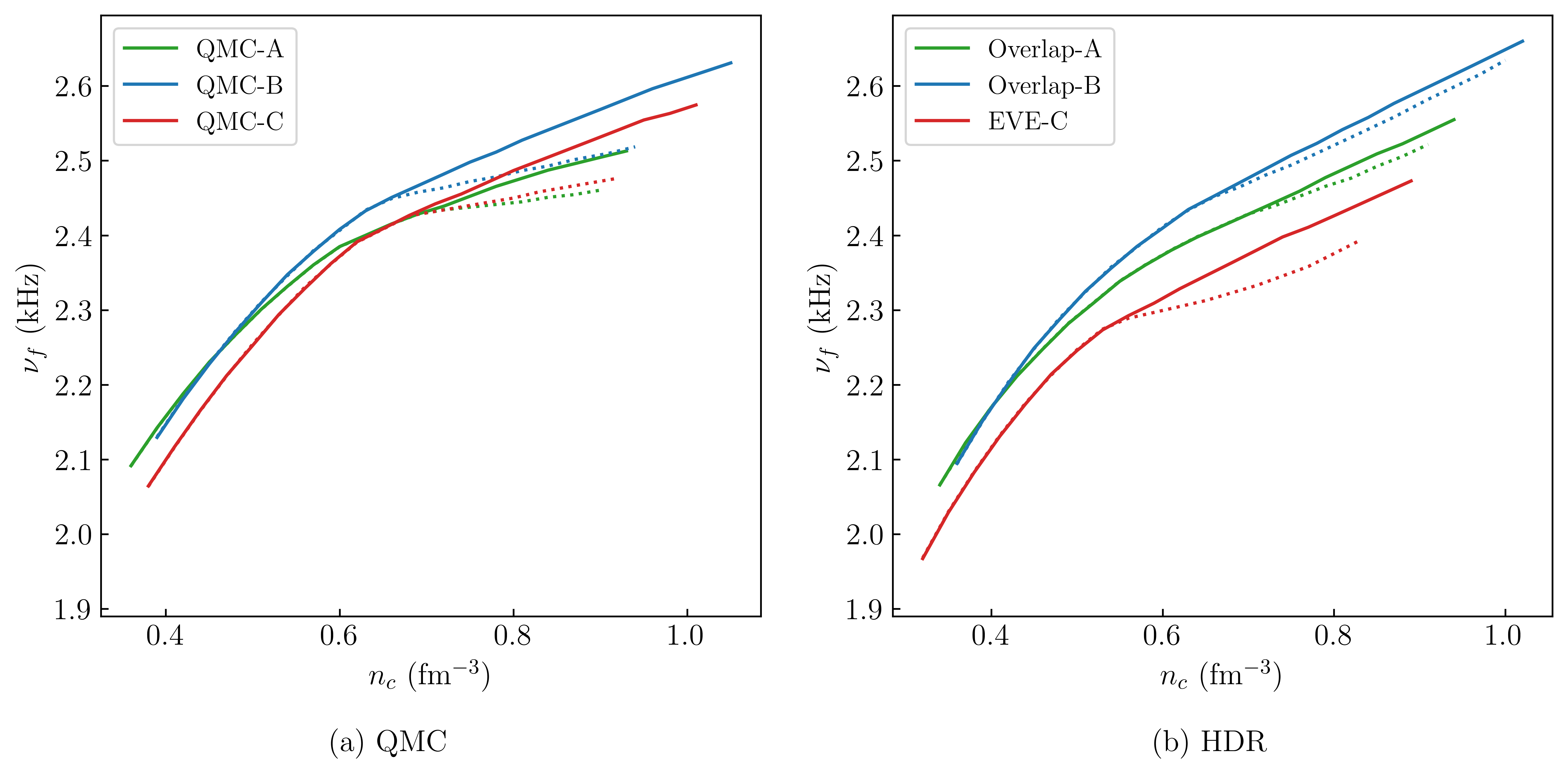}
    \caption{The central number density, $n_c$, for each of the masses presented in 
    Fig.~\ref{fig:MR} and their relationship with the $f$-mode frequencies, $\nu_f$. Near maximum mass, the difference in $f$-mode frequencies between the cases with and without the H-particle are a result of the increase in $n_c$. Once again, the dotted curves indicate that the H-dibaryon has been included.}
    \label{fig:fdens}
\end{figure*}

One can check whether the H yields a stiffer EoS at higher densities when it replaces the $\Xi^0$. The mass of the NS increases from $1.0-1.8M_\odot$, without significant changes to the radius. Beyond this point, small increases in mass will increase the self-gravity further and decrease the radius, pushing it closer to the Schwarzchild limit. This happens sooner for the softer H EoS, as seen by the bifurcation points in Fig.~\ref{fig:MR}, which indicate the presence of the H particle. Leong {\em et al.}~\cite{Leong:2023yma} showed that between the near maximum and maximum mass there is a dramatic increase in the central number density ($n_c$), energy density ($\epsilon_c$) and pressure ($p_c$). This provides the opportunity for an initially soft EoS to become considerably stiffer at high density.

The speed of sound, $c_s^2=\frac{dp}{d\epsilon}$, can be used as a measure of the softness of the EoS. 
Table~\ref{tb:central} shows the central properties of the star at maximum mass for HDR Overlap-B and EVE-C. The corresponding $f$-mode frequency and $c_s^2$ are also provided. The table shows that at a fixed mass of $2.09M_\odot$, which is the maximum for the Overlap-B set with H, $\nu_f=2.64$ kHz is higher than in the case without the H, $\nu_f=2.55$ kHz, even though the former has larger $c_s^2$. As mentioned earlier, the larger speed of sound is attributed to a significant rise in the central densities near and at maximum mass. When comparing these cases at their respective maximum masses, namely $2.09M_\odot$ with H and $2.12M_\odot$ without, the stiffer EoS is Overlap-B without H and it has a higher $\nu_f$. 

Similar results can be seen in the EVE-C set with and without H. For both HDR EoS with the H particle, the $\Xi^0$ was not present. For reference, Fig.~\ref{fig:particle_fraction} shows that for Overlap-B, the $\Xi^0$ threshold is near 1.0 $fm^{-3}$, which is above the density found in a star of mass $2.09 M_\odot$ but below the density at maximum. For EVE, the $\Xi^0$ appears slightly above $0.7$ fm$^{-3}$, so that it is not present in a $1.97 M_\odot$ star but is there at maximum mass. Without the H particle displacing the neutral cascade, the EoS softens and the $f$-mode frequency is increased when the $\Xi^0$ does appear. These results indicate that the suggestion that additional degrees of freedom increase the $f$-mode frequency because of the softening of the EoS does not always hold near maximum mass. 

In Fig.~\ref{fig:fdens} we display the relation between $n_c$ and the $f$-mode frequency for all of the EoS used in this work. At maximum mass, larger central densities yield higher $\nu_f$. This applies to both QMC and HDR EoS, and to the A, B and C parameter sets, which differ from the choice of $m_\sigma$, $\lambda_3$, and NMP used to fix the QMC nucleon-meson couplings.

\begin{figure}
 \centering
\includegraphics[width=1\linewidth]{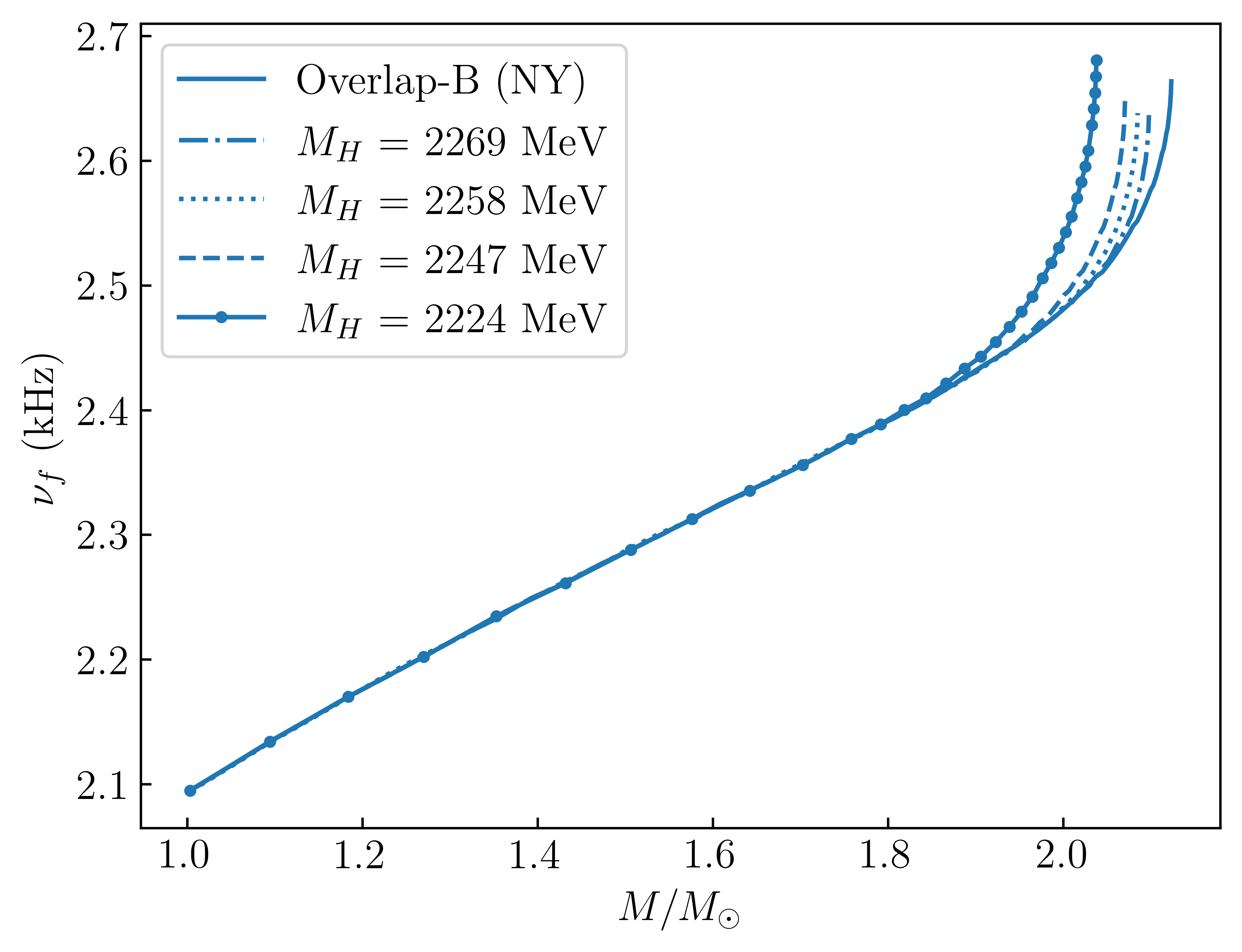}
\caption{The sensitivity of the $f$-mode frequency to variations in the H-dibaryon rest mass ($M_H$) is shown as a function of the NS mass for the Overlap-B EoS parametrization. `NY' denotes the case without H, while the other four cases include the H-particle with different choices of its mass, $M_H$.} 
\label{fig:M_H_var}
\end{figure}

Clearly the H threshold is dependent on its mass in free space and it may even replace the $\Xi^-$ for $M_H=2247$ 
MeV~\cite{Leong:2025fde}. In Ref.~\cite{Leong:2025fde} the authors note that larger values of $M_H$ lead to a larger radius at maximum mass, which should correspond to a smaller $f$-mode frequency. Since in QMC the H only involves one new parameter (its mass), other nuclear models may find different results depending on how the interactions are modelled. It is clearly important to perform a sensitivity study of the $f$-mode frequency with respect to variations in $M_H$. In Fig.~\ref{fig:M_H_var} we display the $f$-mode frequency as a function of neutron star mass for four different choices of $M_H$ consistent with experimental constraints, assuming the Overlap-B EoS~\cite{Leong:2025fde}. The case $M_H = 2258\pm11$ MeV is included in the figure, along with the lower bound from the NAGARA event ($M_H = 2224$ MeV). The relative difference in the $f$-mode frequency between the cases with and without H-dibaryon can vary from $2.5-7.0\%$ near the corresponding maximum mass, because of this uncertainty in $M_H$. The impact is comparatively small for the lattice QCD uncertainty range, whereas adopting the NAGARA lower limit leads to a larger $\nu_f$. For $M_H=2224$ MeV, the only other hyperon present is the $\Lambda$, as the H displaces both $\Xi$ hyperons.

We do find similar results to those just presented when comparing nucleon only and hyperonic EoS at their respective NS maximum mass. In the Appendix~\ref{appendix_sec:fmode_comp_RMF_vs_QMC}, we show a direct comparison between nucleon only EoS and those with hyperons. As discussed above, the $f$-mode frequencies at fixed masses are higher for the cases where hyperons are included. However, our results again differ from those reported in Refs.~\cite{Pradhan_2021} and \cite{Rather_2025} when comparing $\nu_f$ at their maximum masses. There are many differences between the QMC model and the models used in other studies. This is briefly discussed in the Appendix~\ref{appendix_sec:fmode_comp_RMF_vs_QMC} but, just like the H, hyperonic signatures in $\nu_f$ are also likely model-dependent. This could also give us clues to the types of interactions one might expect between baryons at high densities. 

To conclude, we find that the H-included EoS $f$-mode frequencies at maximum mass are consistently lower in both QMC and HDR parametrisations, when compared to hyperonic EoS at their respective maximum masses. Because of this, it is plausible that, although the H is compatible with heavy NS observations, the fundamental frequencies could be used to further distinguish whether a NS has an H-dibaryon condensate within its core. This may carry some model dependence on how the H is parametrized. 
\begin{table*}[htbp]
    \centering
    \begin{tabular}    {|c|ccccc|ccccc|}
    \hline
    Mass & $n_c$ & $p_c$ & $\epsilon_c$ & $c_s^2$& $\nu_f$ &$n_c$ & $p_c$ & $\epsilon_c$ & $c_s^2$ & $\nu_f$\\ 
    ($M_\odot$)& (fm$^{-3}$) & (MeV.fm$^{-3}$) & (MeV.fm$^{-3}$) & & (kHz) &  (fm$^{-3}$) & (MeV.fm$^{-3}$) & (MeV.fm$^{-3}$) & & (kHz) \\ \hline
         & \multicolumn{5}{c|}{Overlap-B with H} & \multicolumn{5}{c|}{Overlap-B without H} \\ \hline
        $2.09^*$& 1.01 & 462 & 1259 & 0.61 & 2.64 & 0.83 & 327 & 973 & 0.59 & 2.55 \\
        $2.12^+$ & \multicolumn{5}{c|}{-}  &1.03& 535& 1307&0.63 & 2.66  \\ \hline
         & \multicolumn{5}{c|}{EVE-C with H} & \multicolumn{5}{c|}{EVE-C without H} \\ \hline
        $1.97^*$& 0.84& 224&981 &0.42 & 2.40& 0.62 & 160 & 678 & 0.36 & 2.33\\
        $2.07^+$& \multicolumn{5}{c|}{-} & 0.91&324 &1105 &0.47 & 2.48\\ \hline
    \end{tabular}
    \caption{Microscopic central properties of maximum mass NS for compatible Overlap-B and EVE-C EoS. $^*$signifies the maximum mass with the H particle. $^+$denotes the maximum mass without the H particle.}
    \label{tb:central}
\end{table*}
%

\subsection{Universal relations}
\label{sec:URs}
GW may carry information from multiple modes simultaneously. The goal of asteroseismology is to deduce some property of the source: mass, radius or moment of inertia, from the detection of GW. This has become known as the inverse problem. A key tool would be finding simple relations between the frequency of the QNM and the macroscopic properties of the NS emitting them. The first attempt was founded upon the theorised relationship between the star's average density and the frequency, as the mechanical perturbations propagate with local sound speed, which is expressed in Eq.~\ref{eq:URdens}. We note that different units for ``$a$" and ``$b$" are used throughout the literature. Here we use dimensionless quantitities, by normalizing the mass and radius as $\bar{M}=\frac{M}{1.4M_\odot}$ and $\bar{R}=\frac{R}{10\text{ km}}$, respectively. Therefore, the units of ``$a$" and ``$b$" are the same as the frequency ($\nu_f$), which is ``kHz". 

The following UR figures separate the 12 EoS (A through to C) into QMC with (green) and without (purple) HDR (Overlap and EVE). All the following UR are calculated from the HDR EoS, with and without H, which reach a maximum mass exceeding $2M_\odot$. The NS mass range is between $1.0M_\odot$ to maximum with no unstable black hole regions included in the fitting procedure. The errors in the parameters are calculated as the square roots of the diagonal elements of the covariance matrix, which is obtained by inverting the normal equation matrix and scaling it by the variance of the residuals. 
\begin{figure}
    \centering
    \includegraphics[width=1\linewidth]{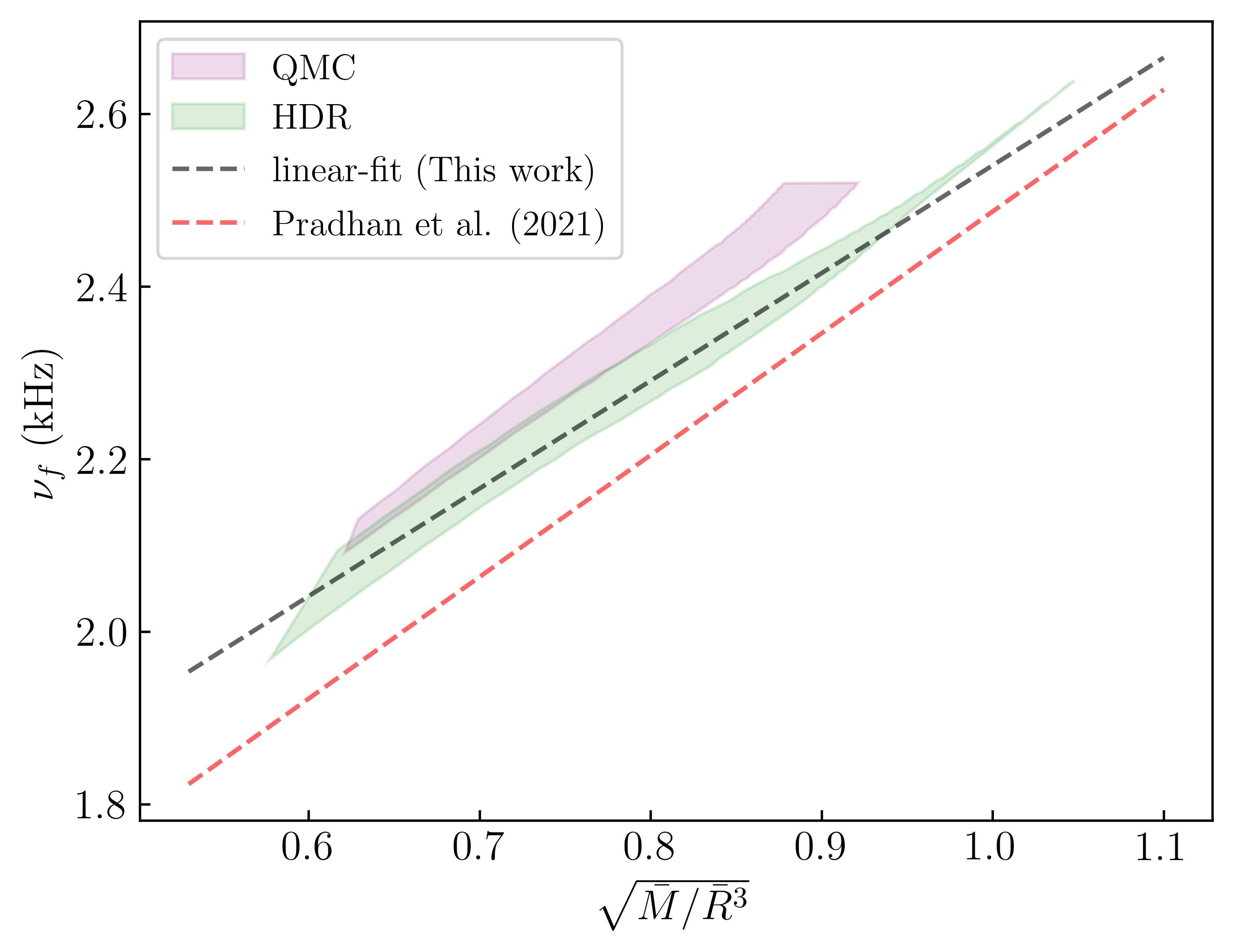}
    \caption{The $f$-mode frequencies are shown as a function of the average density of the star. The purple and green bands represent different parametrisations within the QMC and HDR equations of state, respectively. The linear fit (black dashed line) to Eq.~\ref{eq:URdens} is shown for HDR EoS only, as they are compatible with current observations. The coefficients are presented in Table~\ref{tb:fitdens}.}
    \label{fig:URdens}
\end{figure}
\begin{equation}
\label{eq:URdens}
    \nu_f\;(\text{kHz})= a + b\sqrt{\frac{\bar{M}}{\bar{R}^3}}\ .
\end{equation}

In Fig.~\ref{fig:URdens} we see that there is a quasi-linear relationship between the average density of the star and $\nu_f$. The line of best fit shown in the graph is given by Eq.~\ref{eq:URdens}. The fit parameters are provided in Table~\ref{tb:fitdens} along with other recent studies extending the work first proposed by Andersson and Kokkotas~\cite{Andersson_Kokkotas_MNRAS_1998}. There is a clear separation between the $ f$-mode frequencies with and without HDR. In addition the fitting coefficients across studies are largely different, thus this fit relation is largely EoS-dependent.
\begin{table}[!h]
    \centering
     \resizebox{1 \linewidth}{!}{
    \begin{tabular}{|c|c|c|}
    \hline
         References & $a\;(\text{kHz}$) & $b\;(\text{kHz}$)  \\ \hline 
         Pradhan \& Chatterjee (2021)~\cite{Pradhan_2021} &  $1.075$ & $1.412$  \\
         Rather \textit{et al.} (2025)~\cite{Rather_2025} & $1.29$  & $1.22$ \\
         This work &  $1.29\pm0.01$ & $1.25\pm0.01$ \\
         \hline
    \end{tabular}
    }
    \caption{The coefficients ``$a$" and ``$b$" of the linear fit relation presented by Eq.~(\ref{eq:URdens}).}
    \label{tb:fitdens}
\end{table}

Studies have shown that there is a high degree of correlation between the $f$-mode frequency, $R_{1.4M_\odot}$ and $R_{2M_\odot}$~\cite{Pradhan_2022_fullGR, Shirke_2024_PRD}. This is not surprising, since both depend on the EoS, and the Cowling approximation is solved simultaneously with the TOV equation. Since the $f$-modes are heavily correlated with the structural properties of the stars, UR involving the mass-scaled frequency have been explored. For example, in Eq.~\ref{eq:UR_omegaM_comp} we show the angular frequency, $\omega=2\pi\nu_f$, multiplied by the mass of the star, as a function of the compactness, $C=\frac{M}{R}$. 
\begin{equation}
\label{eq:UR_omegaM_comp}
    \omega M \;(\text{kHz km}) = a \left(\frac{M}{R}\right) + b
\end{equation}

\begin{figure}[!h]
    \centering
    \includegraphics[width=1\linewidth]{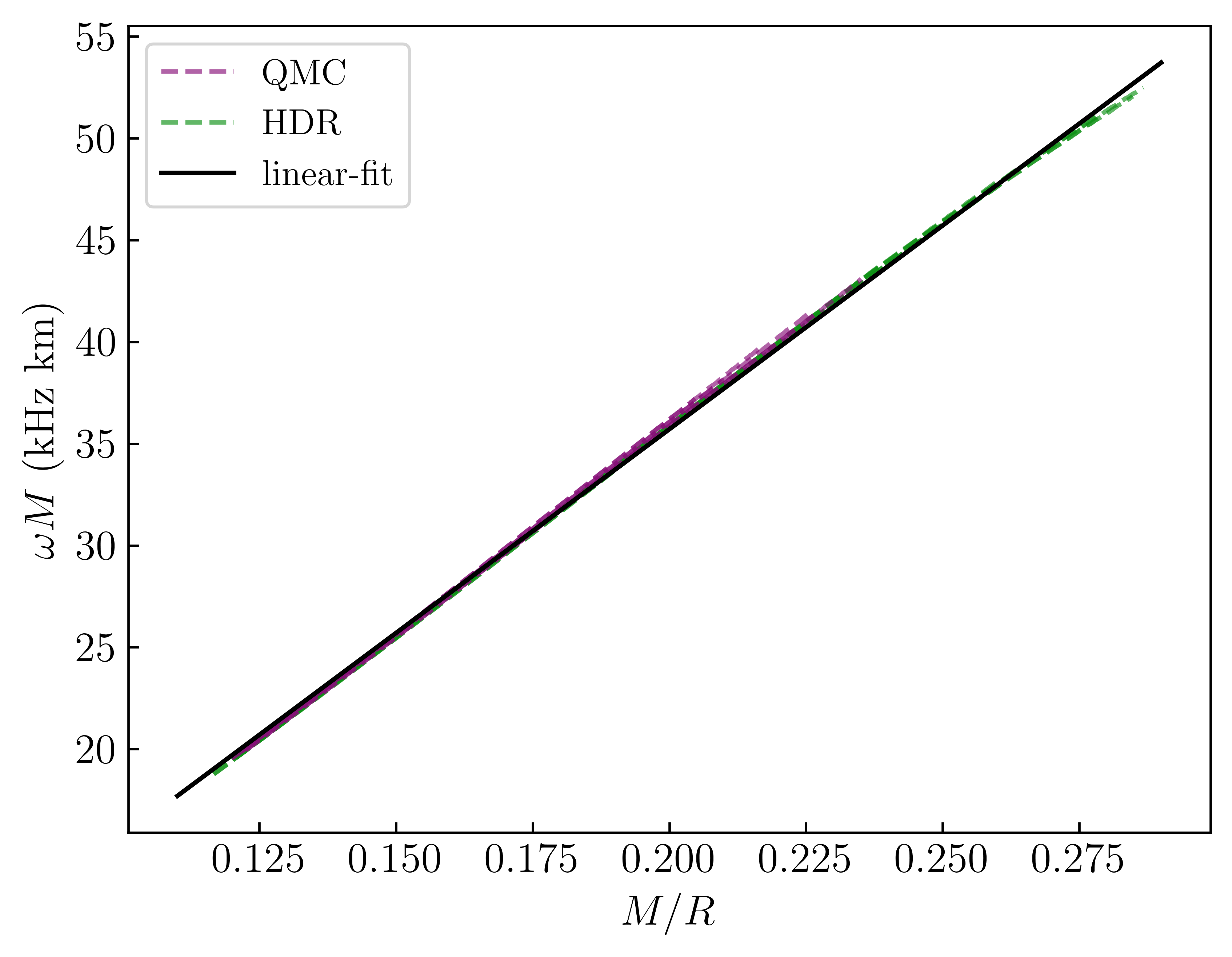}
    \caption{Mass scaled frequencies $\omega M$ as a function of the compactness $M/R$ of the star. HDR EoS are used to fit Eq.~\ref{eq:UR_omegaM_comp} shown in black.}
    \label{fig:UR_omegaM_comp}
\end{figure}

In Fig.~\ref{fig:UR_omegaM_comp}, the mass-scaled angular frequency is plotted against dimensionless compactness,  along with the line of best fit using Eq.~\ref{eq:UR_omegaM_comp}, with $a=200.10\pm0.27$ (kHz km) and $b=-4.31\pm0.06$ (kHz km). One can observe that this relation is quite insensitive to the EoS, as suggested in previous studies~\cite{Pradhan_2021, Rather_2025}. We have also verified that the radius-scaled angular frequency exhibits a quasi-universal relation with compactness, which begins to deviate from universality with increasing mass, as reported in the literature~\cite{Pradhan_2021,Rather_2025}.

In Table~\ref{tb:UR_omegaM_comp}, we present the coefficients of the universal relation shown in Eq.~\ref{eq:UR_omegaM_comp}, along with studies using different nuclear models. It is difficult to compare across studies, since the fits depend on the range of NS masses included. Previous studies have stated that the EoS is constrained by the requirement that it yields a maximum mass greater than $2.0M_\odot$ but have included EoS which generate NS with masses in excess of $2.4-2.5M_\odot$~\cite{Pradhan_2021}. Only one case, PSR J0952-0607, has been detected with such a high mass but large and model dependent uncertainties place the lower limit in this case close to $2.1M_\odot$~\cite{Bassa:2017zpe, Romani:2022jhd}. It is also unclear whether unstable regions of the star have been included in the previous fits in the studies reported in Refs.~\cite{Pradhan_2021, Rather_2025, Dey_2025}. The high NS mass regions do start to deviate from the line of best fit and thus outliers will affect the slope ``$a$" by decreasing it if these high mass regions are either unphysical or over-presented. Because of this, it is expected that the UR is less reliable at masses near maximum.
\begin{table}[!h]
    \centering
    \resizebox{1 \linewidth}{!}{
    \begin{tabular}{|c|c|c|}
    \hline
         References & $a\;(\text{kHz km}$) & $b\;(\text{kHz km}$)  \\ \hline 
         Pradhan \& Chatterjee (2021)~\cite{Pradhan_2021} &  $197.295$ & $-3.836$  \\
         Rather \textit{et al.} (2025)~\cite{Rather_2025} & $199.40$ & $-3.66$ \\
         This work & $200.10\pm0.27$ & $-4.31\pm0.06$ 
        \\ \hline
    \end{tabular}
    }
    \caption{The coefficients ``$a$" and ``$b$" of the linear fit relation presented by eqn.~(\ref{eq:UR_omegaM_comp}).}
    \label{tb:UR_omegaM_comp}
\end{table}
\begin{figure}
    \centering
    \includegraphics[width=1\linewidth]{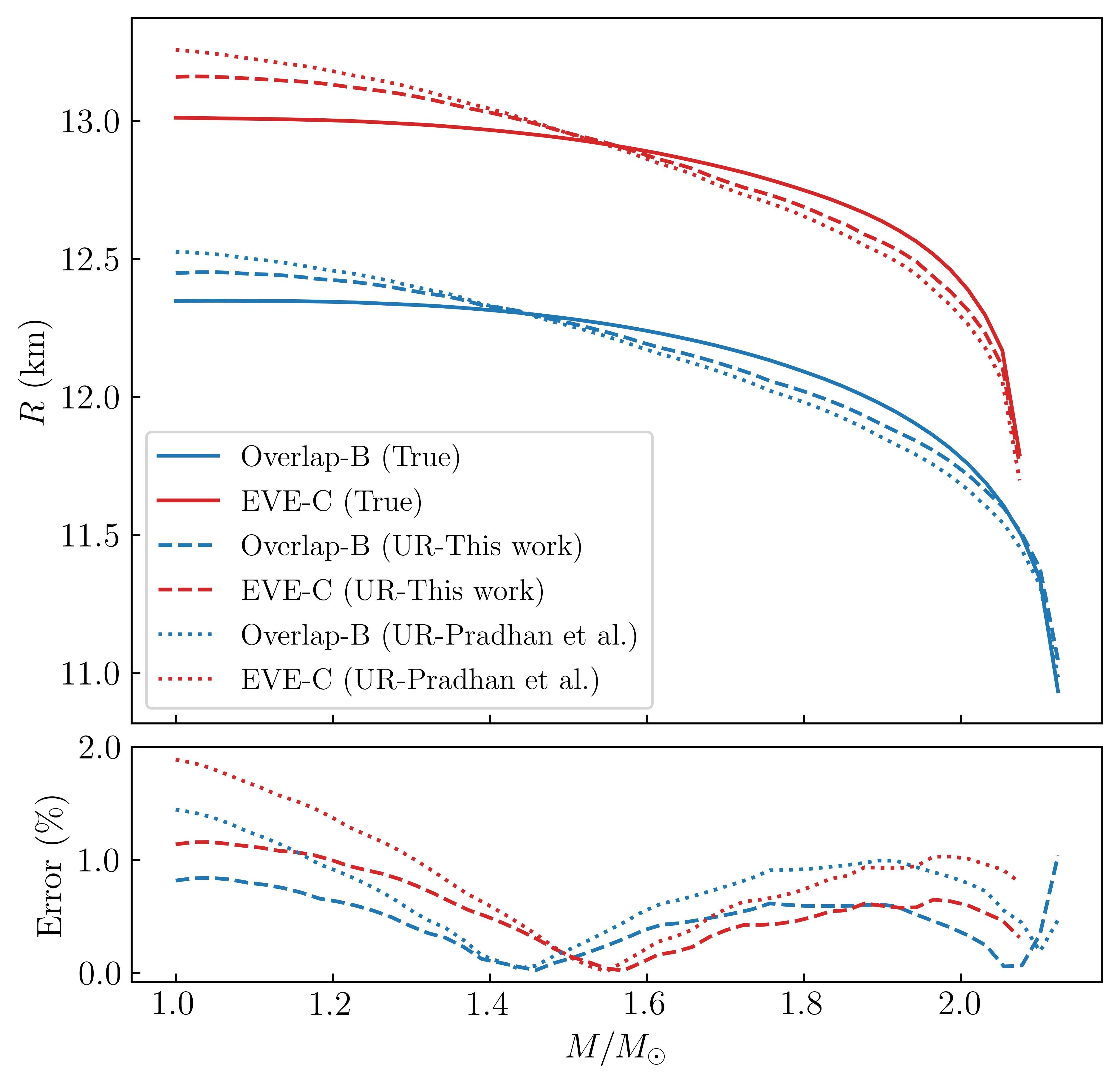}
    \caption{We examine the validity of the mass-scaled frequency UR with compactness using Eq.~\ref{eq:UR_omegaM_comp}. The fit coefficients are given in Table~\ref{tb:UR_omegaM_comp} for our work and reference~\cite{Pradhan_2021}. The mass and frequencies are extracted from Overlap-B and EVE-C, with the radius calculated in the top panel. The bottom panel shows the relative error with respect to the true values.}
    \label{fig:crossref}
\end{figure}

Bearing these concerns in mind, we can ask how similar the fit relations from the different studies presented in Table~\ref{tb:UR_omegaM_comp} are, and what effect any differences may have in inverse asteroseismology. The UR reported in Pradhan \& Chatterjee~\cite{Pradhan_2021} included parametrised EoS with nucleons only and hyperons with varying potentials. Work presented by~\cite{Rather_2025} included nucleon only, hyperonic, $\Delta$, and phase transitions to deconfined quark matter. Both studies display NS maximum mass with ranges from $2-2.5M_\odot$. In Fig.~\ref{fig:crossref} we display the radius deduced from the relation presented in Eq.~\ref{eq:UR_omegaM_comp}, with mass and frequency values chosen from Overlap-B and EVE-C without the H-particle; two EoS parametrisations (HDR) which are known to fit current NS mass, radii and tidal deformability constraints~\cite{Leong:2023yma, Leong:2023lmw}. We also include the results of Ref.~\cite{Pradhan_2021}, where the same relation was fitted in the Cowling approximation, with and without hyperons. All values for the coefficients are provided in Table~\ref{tb:UR_omegaM_comp}. 

In all cases, the accuracy of the estimate of the radius of the star using the UR is better for canonical mass stars, when compared to the low or high mass regions. There appears to be a small bias in overestimating the radius in the low mass region and slightly underestimating the radius in the high mass region except at maximum mass. The errors in radius estimation associated with EVE-C are comparatively larger than those for Overlap-B. This is considered small and due to the sampling of EVE-C EoS with a higher $R_{1.4}$. The UR reported in Pradhan \textit{et. al}~\cite{Pradhan_2021} displays slightly larger errors than ours for all masses except for Overlap-B at maximum. Overall, both works have an error less than $2\%$ across the mass range of the star. Current mass radius measurements carry larger uncertainty than this. Thus one can conclude that the $\omega M$ vs $\frac{M}{R}$ UR is similar across studies.

\begin{equation}
\label{eq:URtidal}
    \omega M \;(\text{kHz km}) = \sum_i \alpha_i [ln({\Lambda})]^i.
\end{equation}
\begin{table*}[htbp]
    \centering
    \begin{tabular}{|c|c|c|c|c|}
    \hline
       Fitting types  & $\alpha_0$(kHz km) & $\alpha_1$(kHz km) & $\alpha_2$(kHz km) & $\alpha_3$(kHz km)  \\ \hline 

         linear-fit & $65.02\pm0.04$ & $-5.83\pm0.01$ & - & - \\ 
         cubic-fit & $60.48\pm0.20$ & $-2.31\pm0.14$ & $-0.83\pm0.03$ & $0.06\pm0.002$\\
         \hline
    \end{tabular}
    \caption{The coefficients ($\alpha_i's$) of the expansion shown in eqn.~(\ref{eq:URtidal}) for the `$f$-Love' universal relation.}
    \label{tb:URtidal}
\end{table*}
\begin{figure}[!h]
    \centering
    \includegraphics[width=1\linewidth]{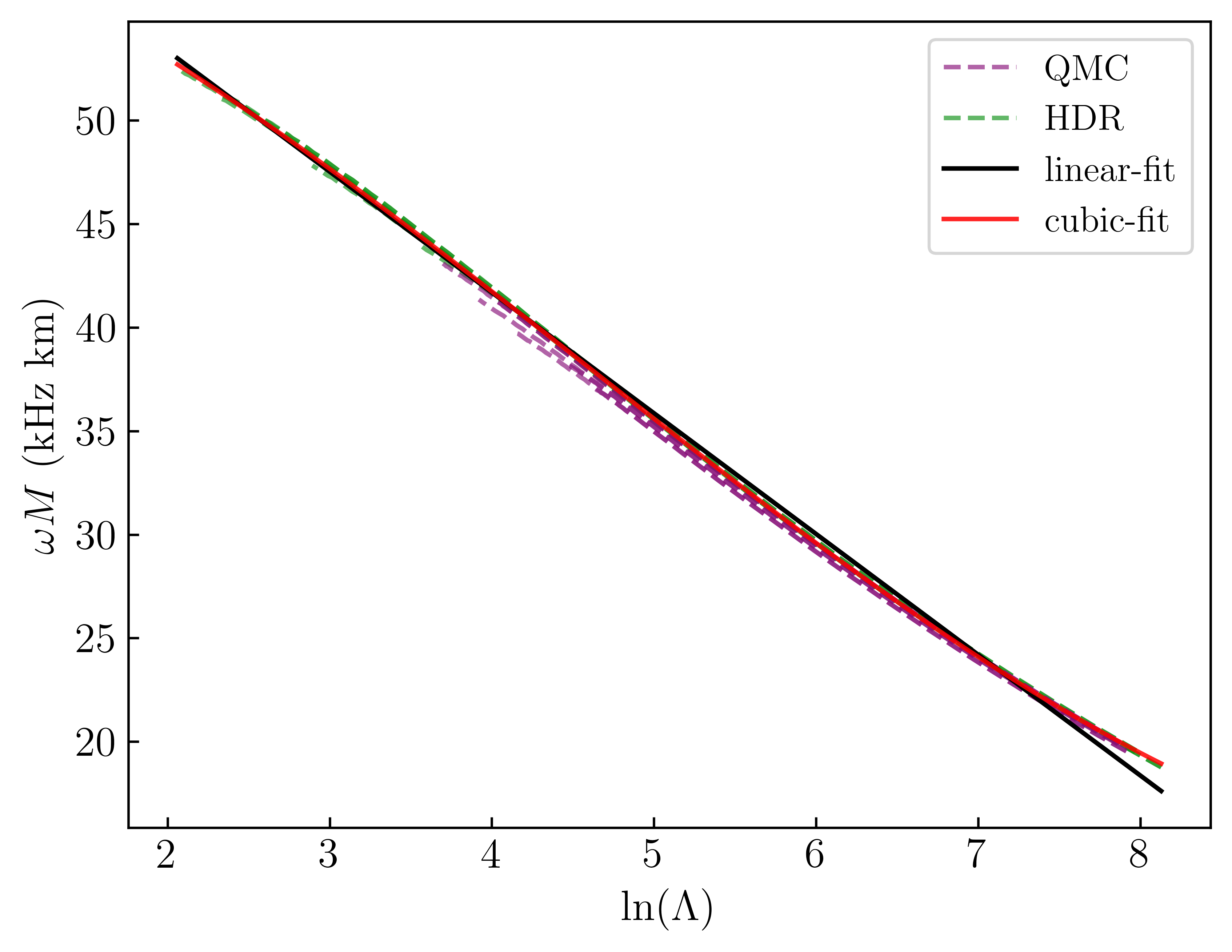}
    \caption{$\omega M$~as a function of the logarithmic tidal deformability (ln$(\Lambda)$) of the neutron stars for the EoS parametrisations used in this work. The provided fits accord with Eq.~\ref{eq:URtidal}, considering HDR EoS only.} 
    \label{fig:URtidal}
\end{figure}
Beside the compactness, tidal deformability and moment of inertia have been shown to obey UR for the QNM. The relation for tidal deformability stated in Eq.~\ref{eq:URtidal} is especially promising, because it is theoretically possible to extract both the fundamental frequency and the tidal deformability from a single GW detection event. In Fig.~\ref{fig:URtidal}, we display the mass-scaled frequency as a function of logarithmic tidal deformability for the EoS parametrisation used in this work, along with the best-fit lines. The coefficients for the linear and cubic fits associated with Eq.~\ref{eq:URtidal}, are listed in Table~\ref{tb:URtidal}. The linear fit does well except for large values of ${\Lambda}$ corresponding to low mass stars. 

\begin{figure}[!h]
    \centering
    \includegraphics[width=1\linewidth]{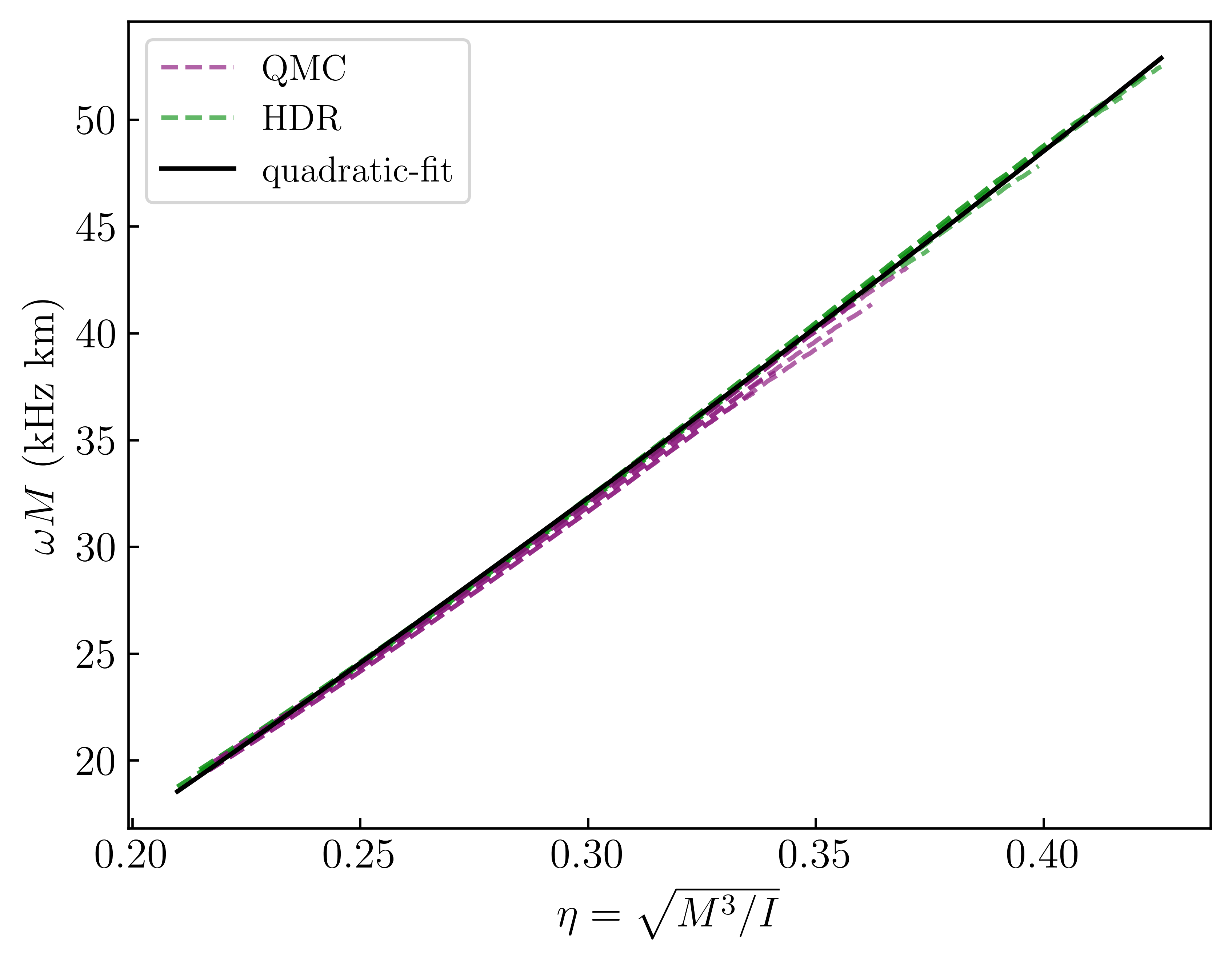}
    \caption{$\omega M$~as a function of the effective compactness,  the dimensionless parameter $\eta$. The quadratic fit for the HDR EoS only is given by Eq.~\ref{eq:UR_MoI}.}
    \label{fig:UR_MoI}
\end{figure}

The final UR relation analysed here is that proposed by Lau \textit{et al.}~\cite{Lau_APJ_2010}, who showed that the mass-scaled frequency, $\omega M$, would be some function of a dimensionless parameter $\eta=\sqrt{\frac{M^3}{I}}$, with $I$ indicating the moment of inertia. The moment of inertia is sensitive to the radial mass distribution. In Fig.~\ref{fig:UR_MoI}, we show $\omega M$ against $\eta$. The solid black line is a quadratic fit following the form given in Eq.~\ref{eq:UR_MoI}, considering the HDR EoS only. A direct comparison with Lau \textit{et al.} is not possible, as our calculation is performed in Cowling approximation, whereas theirs involved full GR. We do find a small difference between QMC with and without HDR but overall this UR holds. 
\begin{equation}
    \omega M \text{ (kHz km)} = -10.21 + 126.1\eta + 51.82\eta^2.
    \label{eq:UR_MoI}
\end{equation}
%


\subsection{Detectability}
\label{sec:detectability}
The potential mechanisms for the excitation of $f$-modes are  glitching of isolated pulsars or perturbations due to the tidal force exerted by the companion during the late inspiral phase in a binary merger. The burst signal from the $f$-mode excitation of a glitching pulsar can be modelled as a damped sinusoidal waveform. Hence, the strain $h(t)$ of the signal having frequency $\nu_f$ and damping time $\tau$ is given by~\cite{Ho_2020}
\begin{equation}
    h(t) = h_0 e^{-t/\tau}\text{sin}(2\pi\nu_f t) \; \;\;\;\;\;\text{for}\; \; t \ge 0 \, .
\end{equation}
The amplitude $h_0$ measured on Earth by the detectors from a source at distance $d$ and associated glitch energy $E_{GW}$ can be written as,
\begin{equation}
    h_0 = 1.53 \times 10^{-17} \sqrt{ \frac{E_{GW}}{M_\odot c^2}} \left( \frac{1\; kpc}{d} \right) \left( \frac{1\; kHz}{\nu_f}\right) \sqrt{\frac{1\;s}{\tau}} \, .
\end{equation}
The signal-to-noise ratio (SNR) for this kind of signal is 
\begin{equation}
\label{eq:SNR}
    SNR = \sqrt{\frac{4Q^2}{1+4Q^2}} h_0 \sqrt{\frac{\tau}{2 S_n}},
\end{equation}
where $Q=\pi \nu_f \tau$ is the quality factor of the signal and $\sqrt{S_n}$ is the noise curve of the detector as a function of frequency. As the frequency of the $f$-mode lies between $1-3$ kHz and the typical damping time is in the order of one tenth of second, $4Q^2 >> 1$, the Eq.~\ref{eq:SNR} is written as, 

\begin{equation}
\label{eq:SNR_approx}
    SNR \approx 1.53 \times 10^{-17} \sqrt{ \frac{E_{GW}}{M_\odot c^2}} \left( \frac{1\; kpc}{d} \right) \left( \frac{1\; kHz}{\nu_f}\right) \sqrt{\frac{1\;s}{2 S_n}} 
\end{equation}

\begin{figure*}[htbp]
    \centering
    \includegraphics[width=1\linewidth]{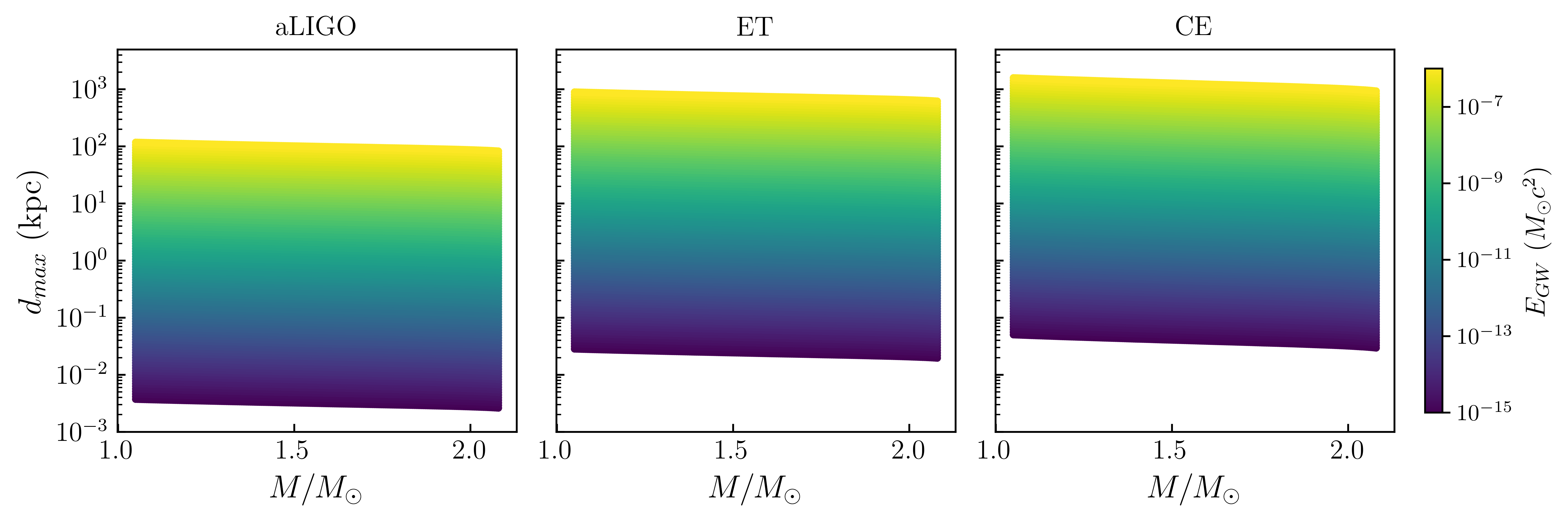}
    \caption{Maximum distances ($d_{max}$) up to which detection would be possible for the detectors aLIGO, ET, and CE are shown as a function of neutron star masses, assuming SNR $\ge$ 5 and the Overlap-B parametrization. The colour bar shows the typical range of glitch energy that excites $f$-modes. The sensitivity curves for aLIGO, ET and CE are obtained from \href{https://dcc.ligo.org/LIGO-T1500293/public}{https://dcc.ligo.org/LIGO-T1500293/public}.}
    \label{fig:detectability}
\end{figure*}

With the rapid advances in the sensitivity of GW detectors one may expect to detect the signal of $f$-mode oscillations from many burst sources using next-generation detectors (ET and CE)~\cite{Maggiore_2020, Hall_2021}. We quantify the maximum distances ($d_{max}$) up to which it will be possible for the detectors to detect signals from the glitching sources of different masses depending on $E_{GW}$ values. Assuming an EoS parametrization, the frequency of the $f$-mode becomes fixed for a neutron star mass. Considering SNR $\ge$ 5 as the detection threshold, we obtain $d_{max}$ as a function of neutron star mass for a typical range of $E_{GW}$ assuming the Overlap-B parametrization, displayed in Fig.~\ref{fig:detectability}. We observe that the future detectors ET and CE should be able to detect up to one order of magnitude greater distance compared to the advanced-LIGO (aLIGO) detector. Using the Cowling-approximation frequency in Eq.~\ref{eq:SNR_approx} introduces a systematic error in the estimation of $d_{max}$, but we expect that the overall order of magnitude of $d_{max}$ will remain unchanged. We have also verified that the EoS uncertainties considered here contribute at most $\sim 20\%$ relative uncertainty in the estimate of $d_{max}$. However, the maximum detectable distance depends strongly on the unknown glitch energy $E_{GW}$.

In order to distinguish between the cases with and without the H-dibaryon through future $f$-mode observations, the detected SNR must be sufficiently high that the error in the frequency measurement is smaller than the reported relative difference in frequency between these two cases. Since the Cowling approximation introduces systematic uncertainties compared to a full GR treatment, we provide an estimate of the difference between the Cowling and GR scenarios in Appendix~\ref{appendix_sec:cowling vs GR} to assess and verify the magnitude of the relative frequency differences between the two cases (with and without H) in both formalism, which is relevant for the future $f$-mode detection. There is another complementary observable, the damping time ($\tau$), which can also be useful to probe the emergence of the H-dibaryon inside massive NS. However, determining the damping time will be more challenging from a GW detection. Within the Cowling approximation, the frequency is purely real and, therefore, the damping time cannot be computed directly. Nevertheless, differences in the damping times between the cases with and without the H-dibaryon are expected for massive NS (see Appendix~\ref{appendix_sec:cowling vs GR}). Alternatively, the tidal deformability can also be used to distinguish between the two cases with and without H for heavier neutron stars near maximum mass. In addition to the variation in damping time, we have checked that the relative difference in the dimensionless tidal deformability between the cases with and without H reaches up to $30\%$ near maximum mass, for the HDR EoS parametrizations.

\section{Conclusion}
\label{sec:conclusion}
Since the pioneering works by Andersson and Kokkotas~\cite{Andersson_Kokkotas_PRL_1996,Andersson_Kokkotas_MNRAS_1998}, NS asteroseismology has been widely explored as an alternative and direct probe of the composition of the interior of neutron stars. There exist a number of works that studied the effects of the NS EoS on oscillation modes, crucial for the detection and correct interpretation of GW data. Several recent works investigated the effect of hyperons in the NS core on the fundamental $f$-mode oscillations using different dense matter nuclear models, e.g. within the framework of the relativistic mean field (RMF) model~\cite{Pradhan_2021,Pradhan_2022_fullGR}, within the density-dependent RMF model framework~\cite{Thapa_2023,Rather_2025} and within the nuclear meta-model description~\cite{Maiti_2024,Montefusco_2024}. These studies included both Cowling approximation and a full GR description, establishing correlations among nuclear and NS astrophysical parameters and providing fits for universal relations compatible with state-of-the-art data. These studies serve to compare the calculations and identify which of the correlations arise from the underlying nuclear model.

For hyperonic matter, the QMC model has been very successful in reproducing nucleonic and hyperonic matter consistently within the same framework. Recently, the model was also extended to include HDR terms to ensure compatibility with large observed NS masses. This work presents the first study of $f$-modes using the QMC model, including hyperons with or without the presence of H-dibaryons, and compares the predictions with the results of previous studies in the literature. UR, which are of considerable interest in asteroseismology, are investigated and the fit parameters compared to other recent works.

The QMC model is fundamentally different from the other hadronic theories, such as the RMF model, in its description of many-body forces and in including Fock terms. While the nucleon-meson couplings are fitted to the saturation properties of nuclear matter such as saturation density, binding energy per nucleon and symmetry energy, non-linear terms such as the cubic scalar self-interaction term ensure compatibility with experimental data for giant monopole resonances; the latter related to the nuclear incompressibility. Including hyperons in the QMC model does not require any additional parameters, while for the H-dibaryon the only additional parameter is its as yet unknown mass. 

In the case of the non-linear RMF model, all the meson-nucleon couplings are fitted to saturation nuclear properties, including incompressibility, effective nucleon mass and slope of symmetry energy. Hyperons are included by introducing hyperon-meson couplings, which are fitted to hypernuclear data. 
The appearance of hyperons as additional degrees of freedom leads to a significant softening of the EoS, resulting in a lower maximum mass and reduced radii. 
A lower NS radius was found to correspond to a higher $f$-mode frequency. In those works, it was therefore concluded that the appearance of hyperons leads to higher $f$-mode frequencies.

A major focus of this investigation to examine the $f$-modes derived from the QMC EoS and to see if the hyperonic and H-dibaryon EoS may be distinguishable. Since the threshold for H-dibaryon is close to that of $\Lambda$, the appearance of H effectively replaces the $\Lambda$ hyperons in the EoS. The s-wave condensate contributes no direct pressure and leads to a slightly softer EoS. When the HDR is utilised, the softening of the EoS is less appreciable. Initially then the H has a higher $f$-mode frequency. But comparison at respective maximum masses between hyperonic and H EoS show that $\nu_f$ is smaller when the H is included. The difference in $\nu_f$ is small, making the differentiation between H and non-H EoS solely using $f$-modes questionable. It is likely to be idiosyncratic to the QMC modelling procedure, since the hyperons and H potentials are all internal to QMC. In combination with the flavour independent HDR, the differences in the $f$-mode frequency between the H and hyperonic EoSs are at most a little over $3\%$. Other RMF models of the H may find more significant differences between H and non-H EoS, as the couplings of the hyperons and H need to be fitted independently of one another, along with possibly a different set of parameters to satisfy $M_{max}>2.0 M_\odot$.

We also investigated the UR related to $f$-mode frequencies and compared them with fits derived in earlier works for hyperonic matter within the Cowling approximation. For the quasi-universal relation of $f$-modes with density, we provided fit coefficients, along with those from earlier studies. For the relation between mass-scaled $f$-mode frequencies with compactness, which has been reported as EoS-independent, the fit coefficients of the QMC model including HDR, with or without H-dibaryons, were found to be compatible with earlier studies. We also demonstrated explicitly that the difference in the estimates of radii using the different fits is less than 2\%. Finally, we also provided useful fit coefficients for the mass-scaled frequencies of $f$-modes as a function of the tidal deformability parameter, $\ln {\Lambda}$, as well as the parameter $\eta = \sqrt{M^3/I}$, where $I$ is the moment of inertia. These fits can be useful in GW asteroseismology in deducing the NS parameters following a successful detection of $f$-modes, independent of the EoS. 

In this work, we employed the relativistic Cowling approximation to estimate the $f$-mode frequencies. This approximation is known to yield errors of up to 30\% compared to a full GR treatment. Furthermore, it does not provide an estimate of the damping timescales. However, in the present study the aim was to test whether the predictions of the QMC model, with HDR and with or without H-dibaryons, concerning $f$-modes were different from those of the non-linear RMF model. For that purpose the conclusions are qualitatively unaffected by assuming the Cowling approximation. 
These studies can easily be extended to a general relativistic formalism, although this is more computationally expensive. 

The study of NS asteroseismology within different nuclear frameworks is important and timely, in order to avoid inaccurate interpretations of GW data. The predicted $f$-mode frequencies are at the limits of sensitivity of the current generation of LVK detectors and there have been no direct detections of these modes. A future generation of detectors, such as the planned ET or CE, are expected to observe many more neutron stars as GW sources, with much higher chances of successful detection of $f$-modes. This will provide important insight into the interior composition of NS and the existence of exotic components such as hyperons and H-dibaryons.

\acknowledgments
This work was supported by the University of Adelaide and the Australian Research Council through a grant to the ARC Centre of Excellence for Dark Matter Particle Physics (CE200100008) and DP230101791 (AWT). The project was initiated during the visit of DC to CSSM in Adelaide, which was supported by the George Southgate Fellowship. DC is thankful to AWT and JL for their warm hospitality at the University of Adelaide. We also thank the Referee whose thoughtful comments helped to further improve the manuscript.

\section*{Data availability}
The data that support the findings of this article are openly available~\cite{DVN/QKTLQI_2026}.


\appendix
\label{sec:appendix}

\section{Model comparison of $f$-mode frequency at maximum mass}
\label{appendix_sec:fmode_comp_RMF_vs_QMC}
\begin{figure}[htbp]
    \centering
    \includegraphics[width=1\linewidth]{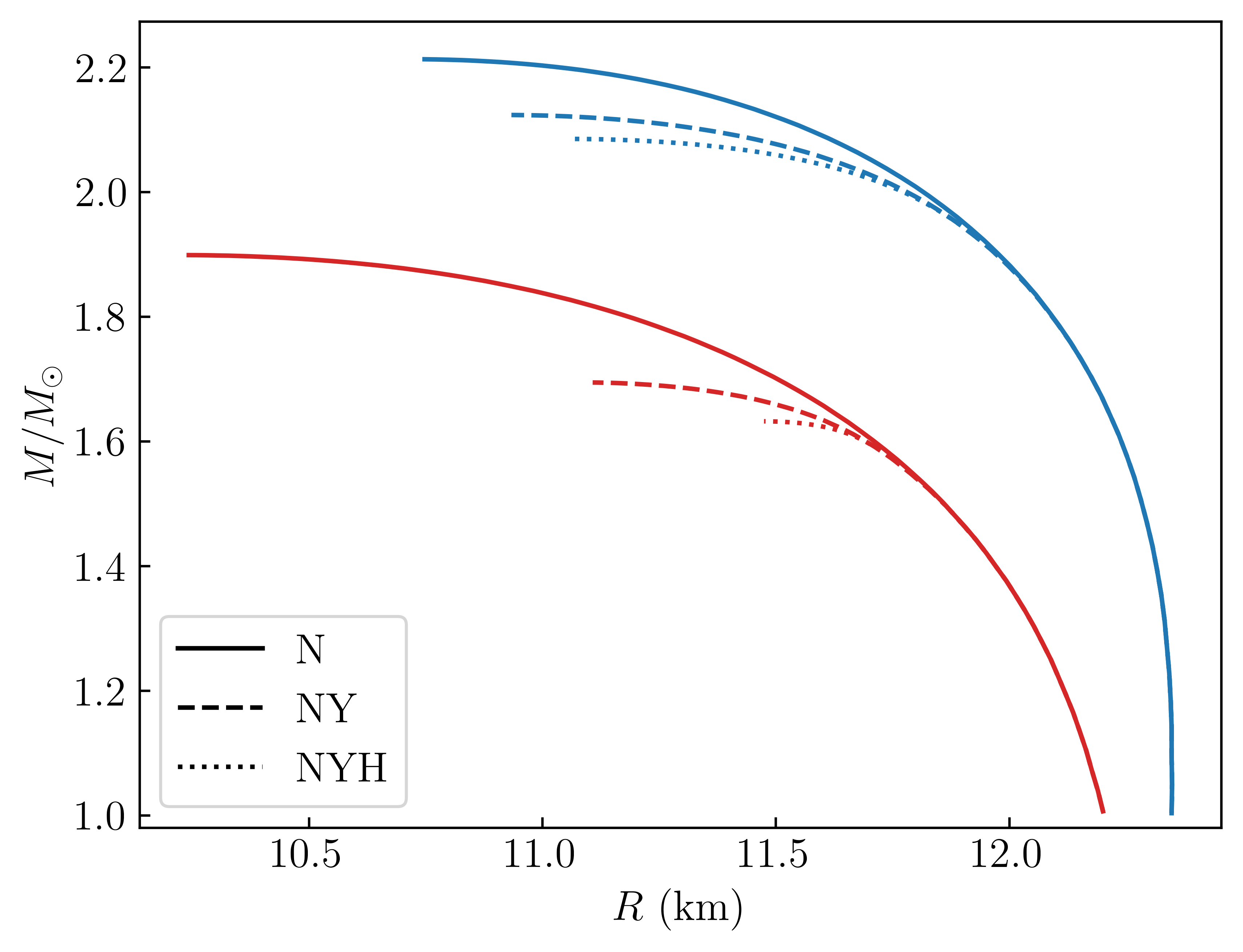}
    \caption{Comparison of mass-radius curves for different EoS containing only nucleons (N), nucleons + hyperons (NY), and nucleons + hyperons + H-dibaryons (NYH). The blue (red) curves represent the Overlap-B (QMC-B) parameter set, as discussed in the Table~\ref{tab:Parameterlist}.}
    \label{fig:mr_comp_N_NY_NYH}
\end{figure}

\begin{figure}[htbp]
    \centering
    \includegraphics[width=1\linewidth]{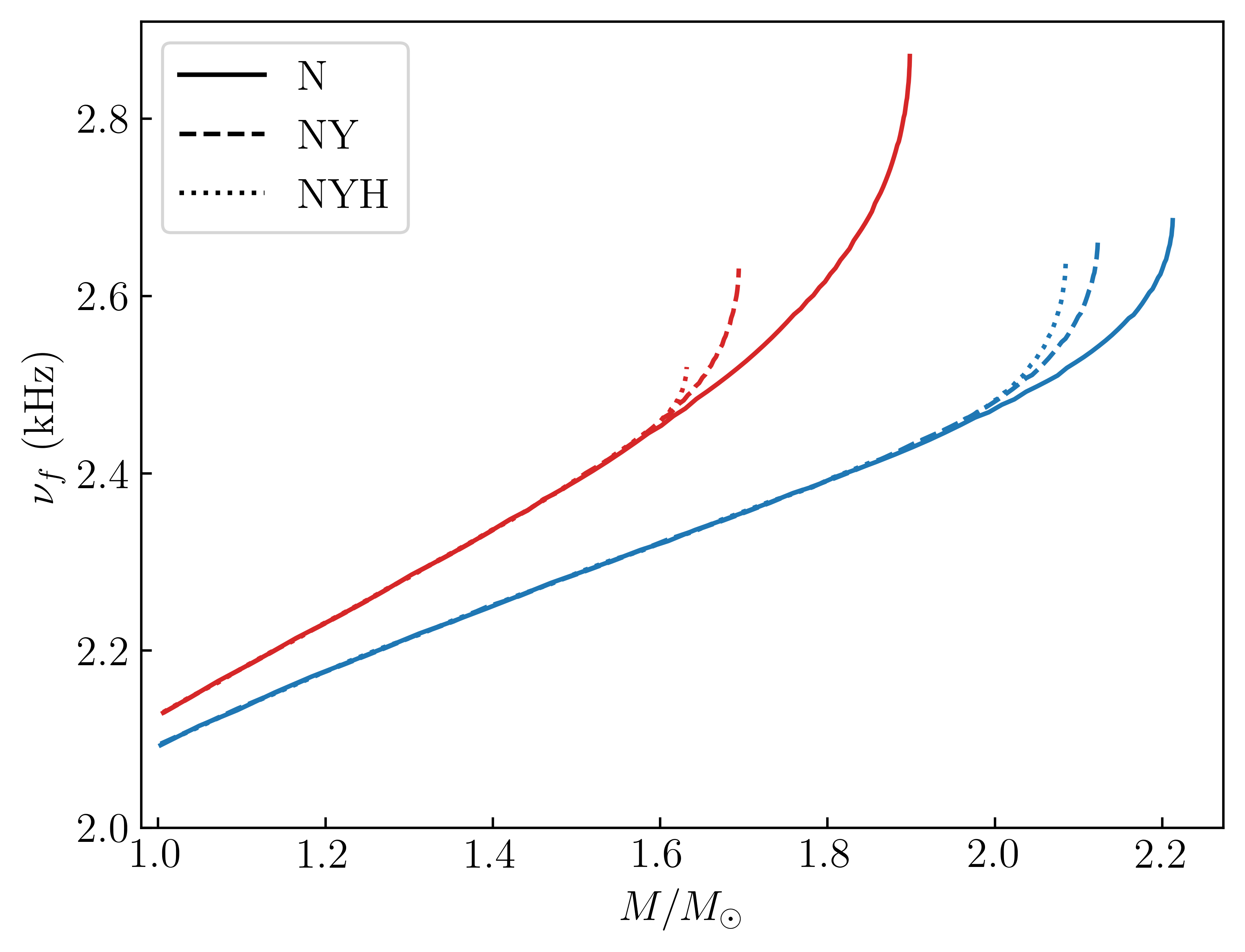}
    \caption{Comparison of $f$-mode frequencies as a function of neutron star mass for different EoS containing only nucleons (N), nucleons + hyperons (NY), and nucleons + hyperons + H-dibaryons (NYH). The blue (red) curves represent the Overlap-B (QMC-B) parameter set, as discussed in the Table~\ref{tab:Parameterlist}.}
    \label{fig:fmode_comp_N_NY_NYH}
\end{figure}

Figure \ref{fig:mr_comp_N_NY_NYH} displays the mass-radius curve for QMC-B and corresponding HDR Overlap-B EoS for nucleonic only, hyperonic and H-dibaryon EoS. These are labelled N, NY, and NYH respectively in both Fig. \ref{fig:mr_comp_N_NY_NYH} and in the resultant $f$-modes calculations displayed in Fig. \ref{fig:fmode_comp_N_NY_NYH}.  
The $f$-mode at maximum NS mass, which we label as $\nu_f^{max}$, is larger for the nucleonic only EoSs, followed by hyperonic EoSs and then NYH. The same pattern holds between QMC-B and Overlap-B, with only a quantitative difference separating them. These results for $\nu_f^{max}$ differ from those presented in both 
Refs.~\cite{Pradhan_2021,Rather_2025}, where the hyperonic EoS had a higher $\nu_f^{max}$ than the nucleon-only case.

As there are many differences between the models used there and here, it is difficult to isolate the origin. However, one outstanding difference is the hyperonic thresholds. While the particle fractions are not shown in 
Refs.~\cite{Pradhan_2021,Rather_2025}, in Rather \textit{et al.} the $c_s^2$ profile indicates that the first hyperon appears around $2n_0$. That is, the hyperons  appear much earlier in these RMF models than in QMC; at a far lower mass than the maximum predicted in the nucleon only case. The hyperons are then able to yield a higher $\nu_f^{max}$ because of the longer interval between their appearance and the maximum possible mass.

The larger hyperon thresholds for QMC are most likely a consequence of the many body interaction presented in 
Eq.~\ref{eq:effM}. The changes in the internal dynamics of the baryon caused by the non-linear term in Eq.~\ref{eq:effM} is weaker for strange baryons. This causes a delay in the appearance of hyperons and is unique to QMC. The thresholds in other RMF models are mainly governed by the coupling strengths of the hyperons to the mesons, which are often free parameters fitted at saturation density to hyperon-nucleon potentials, where known. This difference in hyperonic threshold suggests a degree of dependence on the EoS model, and questions how easily distinguishable the composition of heavy NS are through the examination of quasi-normal modes. 

We have shown that the $\nu_f$ is EoS dependent, but have not ruled out the possibility that $f$-modes can play a vital role in helping to distinguish the NS composition. Because no $f$-modes have yet been detected, the theoretical study of them is still in its infancy. The contrasting $\nu_f^{max}$ reported using QMC, does open up further investigations which could be particularly important in our understanding of the role of hyperons and the high density EoS. Here we have surmised that the many body interaction plays a vital role.

\section{Cowling vs GR comparison}
\label{appendix_sec:cowling vs GR}
\begin{table*}[htbp]
    \centering
    \begin{tabular}    {|c|c|c|c|c|c|c|c|c|}
    \hline
    EoS & $\nu^C_{f}$ & $\nu^{GR}_{f}$ & $\tau$ & $\frac{\nu^{GR}_{f}}{\nu^C_{f}}$ &  $\left(\frac{\nu^C_{f}-\nu^{GR}_{f}}{\nu^C_{f}}\right)$  & $\left|\frac{\Delta \nu_f}{\nu_f}\right|_{C}$ & $\left|\frac{\Delta \nu_f}{\nu_f}\right|_{GR}$ & $\left|\frac{\Delta \tau}{\tau}\right|$\\ 
     parametrizations  & (kHz) & (kHz) & (s) &  & (\%) & (\%) & (\%) &  (\%) \\ \hline
   Overlap-A (without H) & 2.48 & 2.14 & 0.142 & \multirow{2}{2em}{0.86} & \multirow{2}{2em}{14} & \multirow{2}{2em}{1.65} & \multirow{2}{2em}{2.08} & \multirow{2}{2em}{0.54} \\
   Overlap-A (with H) & 2.53 & 2.18 & 0.141 &  &  & & &    \\ \hline

   Overlap-B (without H) & 2.55 & 2.18 & 0.137 & \multirow{2}{2em}{0.86} & \multirow{2}{2em}{14} & \multirow{2}{2em}{3.37} & \multirow{2}{2em}{4.24} & \multirow{2}{2em}{1.44} \\
   Overlap-B (with H) & 2.64 & 2.27 & 0.135 &  & &  & &  \\ \hline

   EVE-C (without H) & 2.33 & 1.93 & 0.154 & \multirow{2}{2em}{0.83} & \multirow{2}{2em}{17} & \multirow{2}{2em}{3.17} & \multirow{2}{2em}{3.53} & \multirow{2}{2em}{4.29} \\ 
   EVE-C (with H) & 2.40 & 2.00 & 0.148 & & &  &  & \\ \hline
    \end{tabular}
    \caption{The $f$-mode frequencies, the relative differences in frequency ($\Delta \nu_f/\nu_f$), the relative differences in damping time ($\Delta \tau/\tau$) between the cases with and without H-dibaryon at the maximum masses of the corresponding H-containing EoS parametrizations ($M_{max,H}$) are shown in GR estimates (using UR) and also compared with the results obtained in the Cowling approximation. Here, $\nu^C_{f}$ and $\nu^{GR}_{f}$ denote the Cowling and GR $f$-mode frequencies, respectively. The values of $M_{max,H}$ for Overlap-A, Overlap-B, and EVE-C are  $2.18M_\odot$, $2.09M_\odot$, and $1.97M_\odot$ respectively. Note that percentage differences are computed using unrounded values.} 
    \label{tb:freq_Cowling_to_GR}
\end{table*}

As discussed earlier, the Cowling approximation introduces $\sim$ $20-30\%$ systematic error in the $f$-mode frequency estimation, with the error reportedly lower nearer the maximum mass~\cite{Rather_2025}. Therefore, we also estimate and provide the corresponding $f$-mode frequencies and the associated damping times (see Table~\ref{tb:freq_Cowling_to_GR}) within GR using the universal fit relations (see Eqs.~\ref{eq:pradhan_UR_GR_freq_comp} and \ref{eq:pradhan_UR_GR_tau_comp}) between the mass-scaled frequency and compactness given in Ref.~\cite{Pradhan_2022_fullGR}. This allows us to avoid solving the full set of perturbed hydrodynamical and metric equations in GR, which would be computationally expensive. We note that the reported errors associated with the coefficients of the UR (in Ref.~\cite{Pradhan_2022_fullGR}) result in maximum relative errors of $\sim$ $0.3\%$ and $0.5\%$ in the frequency and damping time estimations, respectively, for the presented EoS parametrizations. The Cowling-to-GR frequency correction and the relative difference in frequencies and damping times between the cases with and without H-dibaryon at the maximum masses of the corresponding EoS parametrizations with the H particle have also been estimated and reported in GR.

\begin{align}
    \text{Re}(M\omega) &= 0.079C^2 + 0.599C - 0.026 \label{eq:pradhan_UR_GR_freq_comp} \\
    \text{Im}(M\omega) &= 0.09836C^4 - 0.4448C^5 + 0.4915C^6 \label{eq:pradhan_UR_GR_tau_comp}
\end{align}

\newpage
\bibliographystyle{apsrev4-2}
\bibliography{ref}

\end{document}